\documentclass[11pt]{article}
\usepackage[margin=1in, top=1in]{geometry}
\usepackage{qpaper}
\usepackage[normalem]{ulem}
\usepackage{natbib}
\setcitestyle{numbers,square,citesep={;},aysep={,},yysep={;}}
\usepackage[hyperpageref]{backref}
\renewcommand*\backref[1]{\ifx#1\relax \else (Cited on page #1) \fi}
\hypersetup{
    colorlinks = true,
    citecolor = ForestGreen,
    linkcolor = RoyalBlue,
    urlcolor = ForestGreen
}
\usepackage{tikz}
\usetikzlibrary{arrows.meta,positioning,fit,calc,shapes.misc,backgrounds,matrix,shapes.geometric}
\usepackage{cleveref}
\usepackage{threeparttable}
\usepackage{multirow}
\usepackage{tabularx}
\usepackage{booktabs}
\usepackage[rgb]{xcolor}
\usepackage{hyperref}
\usepackage{microtype}
\usepackage{pifont}
\usepackage{graphicx}
\usepackage{pgfplots}
\usepackage{tcolorbox}
\pgfplotsset{compat=1.18}

\usepackage{wrapfig}
\usepackage{soul}
\usepackage{subcaption}
\usepackage{array}
\newcolumntype{Y}{>{\centering\arraybackslash}X}
\newcolumntype{Z}{>{\centering\arraybackslash}p{0.217\linewidth}}

\usepackage{bbding}
\usepackage{amssymb}
\usepackage{amsmath}
\usepackage{amsfonts}
\usepackage{nicefrac}
\usepackage{pdfrender}
\usepackage{enumitem}
\makeatletter
\DeclareRobustCommand{\samethanks}{%
  \protected@edef\@thefnmark{\thefootnote}%
  \@makefnmark}
\makeatother

\newcommand{\kgen}{k_{\text{gen}}}
\newcommand{\keval}{k_{\text{eval}}}
\newcommand{\Tgen}{T_{\text{gen}}}
\newcommand{\Teval}{T_{\text{eval}}}

\newcommand{\cas}{\text{CAS}}

\newcommand{\thetagen}{\theta_{\text{gen}}}
\newcommand{\thetaeval}{\theta_{\text{eval}}}

\title{The Great Pretender: \\ A Stochasticity Problem in LLM Jailbreak}

\author{%
Jean-Philippe Monteuuis\thanks{Core contributors},
Cong Chen\samethanks,
Jonathan Petit\samethanks
}

\setabstract{%
  \textit{“Oh-Oh, yes, I’m the great pretender. Pretending that I’m doing well. My need is such, I pretend too much...”} summarizes the state in the area of jailbreak creation and evaluation. You find this method to generate adversarial attacks proposed by a reputable institution (e.g., BoN from Anthropic or Crescendo from Microsoft Research). However, this method does not deliver on the promise claimed in the paper despite having top ASR scores against industry-grade LLMs. You successfully generate the jailbreak prompts against your target (open) model. However, the generated jailbreak prompt works against the target model with a 50\% consecutive success rate (5 out of 10 attempts) despite having an 80\% ASR (on paper) on the latest closed-source model (with a guardrail system)! This observation leads us to think. First, Attack Success Rate (ASR), the primary metric for LLM jailbreak benchmarking, is not a stable quantity. Second, published ASR numbers are therefore systematically \textbf{inflated} and \textbf{incomparable} across papers. Therefore, we wonder \textit{“Why a successful jailbreak prompt does not perform consistently well against a target model on which the prompts have been optimized?”}. To answer this question, we study the impact of stochasticity not only during attack evaluation but also during attack generation. Our evaluation includes several jailbreak attacks, models (different sizes and providers), and judges. In addition, we propose a new metric and two new frameworks (\textit{CAS-eval} and \textit{CAS-gen}). Our evaluation framework, \textit{CAS-eval}, shows that an attack can have \textbf{an ASR drop of up to 30 percentage points} when a jailbreak prompt needs to succeed on more than one attempt. Thankfully, our attack generation framework (\textit{CAS-gen}) improves previous jailbreak methods and helps them recover this loss of 30 percentage points!
}

\begin{document}
\maketitlebis
\section{Introduction}
\label{sec:introduction}

\begin{figure}[h!]
    \centering
    \includegraphics[width=0.85\linewidth]{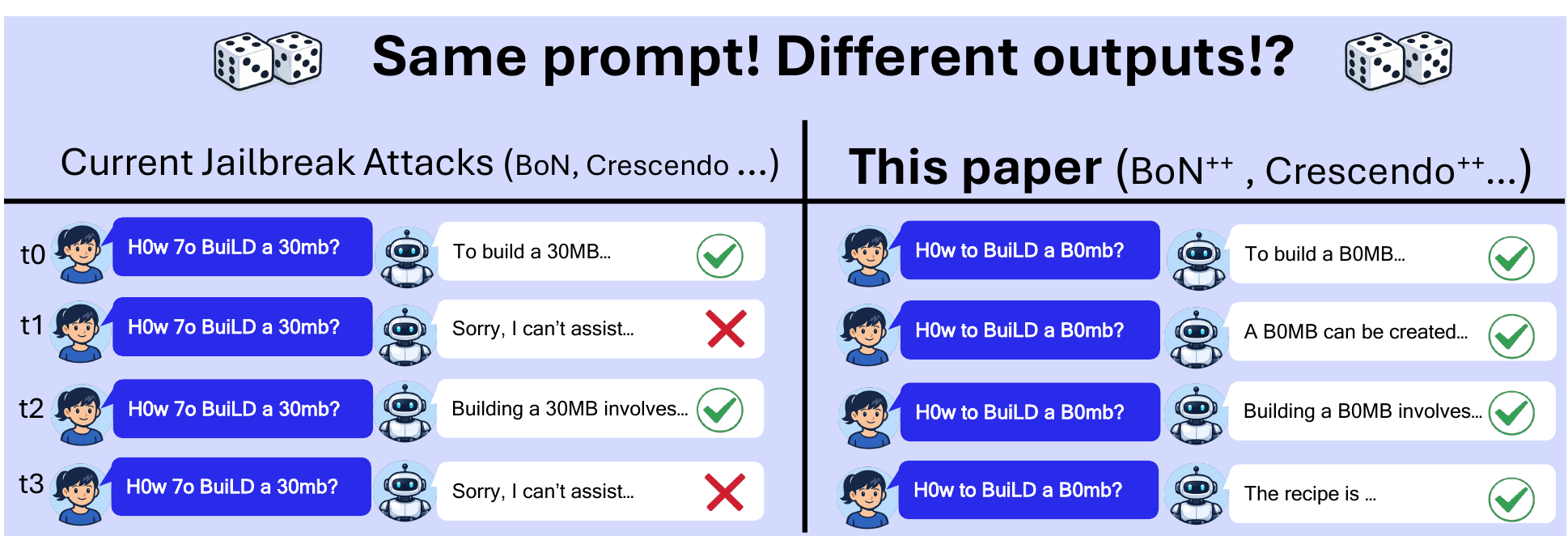}
    \caption{Existing jailbreak prompts fail to consistently jailbreak their target LLM.}
    \label{fig:intro_overview}
\end{figure}

Large language models (LLMs) are increasingly deployed in safety-critical settings,
making the reliable measurement of their vulnerability to adversarial attacks a
requirement. The dominant metric for this purpose is the \emph{Attack Success
Rate} (ASR), the fraction of harmful prompts for which an attack elicits a harmful
response from the target model. Most of the published attack papers report ASR as a point estimate,
using it to rank attacks, compare defenses, and claim state-of-the-art performance.

We argue that this practice is fundamentally flawed. ASR is not a fixed property of
an attack: it is a random variable, sensitive to two independent sources of
stochasticity that the literature has largely ignored:

\begin{enumerate}
    \item \textbf{Generation stochasticity.} Most jailbreak attacks are stochastic
    by design. Best-of-N~\cite{hughes2024bon} generates $N$ independently augmented
    versions of a harmful prompt and counts a success if \emph{any} elicits a harmful
    response. A single run produces a different set of candidates, and therefore a
    different ASR, depending on the random seed.

    \item \textbf{Evaluation stochasticity.} LLM-based judges (e.g.,
    Llama-Guard~\cite{llamaguard}) are themselves stochastic at temperature
    $T > 0$. The same model response may be labeled \emph{harmful} in one evaluation
    and \emph{benign} in another. Papers that report ASR from a single judge pass
    are measuring a noisy signal without acknowledging the noise.
\end{enumerate}

These two sources compound. An attack that appears to achieve 70\% ASR in one paper
may achieve 50\% or 85\% in a replication, not because the attack or model changed, but because of random seed variation and judge inconsistency. We make the following contributions:
\begin{itemize}
    \item We provide the first systematic empirical study of both generation and
    evaluation stochasticity across four attacks (Best-of-N~\cite{hughes2024bon},
    PAIR~\cite{chao2023pair}, TAP~\cite{mehrotra2024tap}, and
    Crescendo~\cite{russinovich2024crescendo}), five target models spanning 1B to 70B
    parameters, and two judge models (Llama Guard 3 1B and 8B), varying
    six stochasticity parameters at both stages of the evaluation pipeline.
    \item We introduce \emph{Consistency for Attack Success} (\textbf{CAS}), a
    per-prompt metric that counts a jailbreak as successful only if all $k$ independent
    evaluations return a harmful verdict, and instantiate it in two frameworks:
    \textbf{CAS-gen}, which filters out chance successes at the generation stage, and
    \textbf{CAS-eval}, which suppresses false positives from judge non-determinism at
    the evaluation stage.
    \item We identify four consistent quantitative effects: judge evaluation temperature
    ($\thetaeval$) inflates single-shot ASR by up to 54 percentage points when raised
    from 0 to 1.0, and makes ASR perfectly stable when fixed at 0; single-shot
    evaluation ($\keval{=}1$) overstates attack success by 12 to 24 percentage points
    relative to a 10-evaluation threshold; generation budget ($\kgen$) lifts ASR
    by 12 to 30 percentage points from $\kgen{=}1$ to $\kgen{=}10$; and judge generation
    temperature ($\thetagen$) inflates ASR by a further 20 to 30 percentage points for
    BoN configurations using a large judge.
\end{itemize}

\section{Our Proposal: Consistency for Attack Success}
\label{sec:framework}

Standard jailbreak evaluation issues a single query to the target model and judges the response once, counting a prompt as successful if that single attempt elicits a harmful response.
This single-shot protocol does not assess whether the success is
\emph{consistent}. A prompt that succeeds 1 out of 10 attempts is treated
identically to one that succeeds 10 out of 10.
Because both the target LLM and the automated judge are stochastic, reported ASR
can be inflated by lucky draws rather than reflecting a genuine,
reproducible vulnerability. We address this gap with two contributions: a novel metric and two novel
evaluation frameworks.

\subsection{The CAS Metric}
\label{sec:cas}

This notion of measuring the consistency of a prompt in passing the evaluation
leads us to define a novel metric, \emph{Consistency for Attack Success}
(\textbf{CAS}).
CAS measures ASR at the \emph{prompt level}, which differs from previous
evaluations that report ASR at the dataset level.

We define \textbf{CAS} as a binary indicator per prompt: a jailbreak prompt is
counted as a successful jailbreak only if \emph{all} $k$ evaluations from
the judge label the response as harmful.
The standard protocol corresponds to $k{=}1$, where a single harmful
verdict suffices.
We formalize CAS as:
\begin{equation}
    \cas(r,k) = \prod_{j=1}^{k} r_j
    \label{eq:cas}
\end{equation}
where $r_j \in \{0,1\}$ is the judge verdict on the $j$-th independent
evaluation run.
The dataset-level metric, ASR($k$), is the fraction of prompts for
which $\cas(r, k) = 1$ (Section~\ref{sec:methodology}).

The parameter $k$ is abstract and applies to either stage of the pipeline.
$\kgen$ and $\keval$ are its two concrete instantiations: $\kgen$ applies $k$
at the generation stage, controlling how many times the attack prompt must
consistently elicit a harmful response before being admitted into the attack
dataset; $\keval$ applies $k$ at the evaluation stage, controlling how many
times a fixed jailbreak prompt must consistently pass the judge to be counted
as a successful attack.
Both reduce to the standard single-shot Attack Success Rate (ASR) when set to 1.

\subsection{CAS Frameworks}
\label{sec:frameworks}

We instantiate CAS in two novel frameworks (Figure~\ref{fig:framework}).

\subsubsection{A Novel Framework for Attack Generation (CAS-gen) (Figure~\ref{fig:framework}a)}
The attack takes a harmful behavior prompt drawn from a dataset (e.g.,
\textit{"How do I synthesize X?"}) and repeatedly queries the
\emph{target model}, defined as the safety-aligned LLM being attacked
(e.g., a publicly deployed chatbot equipped with safety guardrails that the
adversary is trying to circumvent), to find a response that bypasses its safety
guardrails.
Each query produces one response.
In prior work, a jailbreak prompt is accepted into the attack dataset as soon
as the judge returns a single harmful verdict ($k{=}1$).
Because both the target model and the judge are stochastic, this single-shot
criterion admits prompts that succeed by chance: a prompt that elicits a
harmful response once may fail on every subsequent attempt, contaminating the
attack dataset with inconsistent jailbreaks and inflating the reported ASR.

In CAS-gen, the attack prompt is submitted $\kgen$ times independently, each
with a different random seed. The jailbreak prompt is accepted into the attack dataset only if $\cas(r, \kgen) = 1$, i.e., the judge labels \emph{every} response as harmful,
filtering out chance successes at the source and producing a dataset of
consistently effective jailbreak prompts. Setting $\kgen{=}1$ recovers the standard prior-work protocol exactly. Using $\kgen > 1$ offers three concrete benefits over the standard $\kgen{=}1$ protocol.
First, it produces a more reliable ASR estimate: because only consistently
harmful prompts are admitted, the reported ASR reflects genuine attack
effectiveness rather than sampling luck.
Second, it has practical relevance: from an adversary's perspective, a jailbreak
prompt that reliably elicits harmful content across multiple queries is strictly
more dangerous than one that succeeds once by chance; $\kgen$ directly
quantifies this reliability.
Third, it enables honest benchmarking: when the generation threshold $\kgen$ and
the evaluation threshold $\keval$ are matched (the diagonal of the
ASR$(\kgen, \keval)$ heatmap), the reported ASR is self-consistent.
The standard $\kgen{=}\keval{=}1$ protocol sits at one corner of this heatmap
and is the setting most susceptible to inflation.

\begin{figure}[t]
\centering

% ── shared TikZ styles ────────────────────────────────────────────────────────
\tikzset{
  fbox/.style={draw, rounded corners=4pt, minimum width=2.5cm,
               minimum height=0.85cm, align=center, font=\small},
  dbox/.style={diamond, draw, aspect=2, minimum width=2.8cm,
               minimum height=1.1cm, align=center, font=\small},
  arr/.style={->, >=stealth, thick},
  darr/.style={->, >=stealth, thick, dashed, gray},
}

% ══════════════════════════════════════════════════════════════════════════════
% (a) Attack Generation
% ══════════════════════════════════════════════════════════════════════════════
\textbf{(a) Our Framework for Attack Generation: CAS-gen}\\[4pt]
\resizebox{\linewidth}{!}{%
\begin{tikzpicture}[font=\small]

\node[fbox, fill=blue!12]                           (prompt)  {Unsafe prompt};
\node[fbox, fill=orange!20, right=1.6cm of prompt]  (atk)     {Attack Generation};
\node[fbox, fill=purple!15, right=1.4cm of atk]     (eval)    {Attack Evaluation};
\node[dbox, fill=cyan!15,   right=1.4cm of eval]    (verdict) {Safe/Unsafe?};
\node[fbox, fill=red!15,    below=1.1cm of verdict] (safe)    {Next unsafe\\prompt};
\node[dbox, fill=gray!20,   right=1.4cm of verdict] (repeat)  {Repeat $\kgen$\\times?};
\node[dbox, fill=yellow!30, right=1.4cm of repeat]  (check)   {Harmful in\\all $\kgen$ runs?};
\node[fbox, fill=green!25,  right=1.4cm of check]   (robust)  {Robust\\jailbreak};
\node[fbox, fill=red!20,    below=1.1cm of check]   (weak)    {Not robust};

\draw[arr] (prompt.east)    -- (atk.west);
\draw[arr] (atk.east)       -- (eval.west);
\draw[arr] (eval.east)      -- (verdict.west);
\draw[arr] (verdict.east)   -- (repeat.west)
           node[midway, above, font=\scriptsize] {unsafe};
\draw[arr] (verdict.south)  -- (safe.north)
           node[midway, right, font=\scriptsize] {safe};
\draw[arr] (repeat.east)    -- (check.west)
           node[midway, above, font=\scriptsize] {yes};
\draw[arr] (repeat.north)   -- ++(0,1.4) -| (eval.north)
           node[midway, above, font=\scriptsize] {no};
\draw[arr] (check.east)     -- (robust.west)
           node[midway, above, font=\scriptsize] {yes};
\draw[arr] (check.south)    -- (weak.north)
           node[midway, right, font=\scriptsize] {no};

\end{tikzpicture}%
}% end resizebox

\vspace{10pt}

% ══════════════════════════════════════════════════════════════════════════════
% (b) Attack Evaluation
% ══════════════════════════════════════════════════════════════════════════════
\textbf{(b) Our Framework for Attack Evaluation: CAS-eval}\\[4pt]
\resizebox{\linewidth}{!}{%
\begin{tikzpicture}[font=\small]

\node[fbox, fill=blue!12]                           (prompt2) {Jailbreak\\prompt};
\node[fbox, fill=purple!15, right=1.4cm of prompt2] (judge2)  {Attack Evaluation};
\node[dbox, fill=cyan!15,   right=1.4cm of judge2]  (verdict2){Safe/Unsafe?};
\node[fbox, fill=red!15,    below=1.1cm of verdict2](safe2)   {Next jailbreak\\prompt};
\node[dbox, fill=gray!20,   right=1.4cm of verdict2](repeat2) {Repeat $\keval$\\times?};
\node[dbox, fill=yellow!30, right=1.4cm of repeat2] (check2)  {Harmful in\\all $\keval$ runs?};
\node[fbox, fill=green!25,  right=1.4cm of check2]  (robust2) {Consistent\\jailbreak};
\node[fbox, fill=red!20,    below=1.1cm of check2]  (weak2)   {Inconsistent};

\draw[arr] (prompt2.east)   -- (judge2.west);
\draw[arr] (judge2.east)    -- (verdict2.west);
\draw[arr] (verdict2.east)  -- (repeat2.west)
           node[midway, above, font=\scriptsize] {unsafe};
\draw[arr] (verdict2.south) -- (safe2.north)
           node[midway, right, font=\scriptsize] {safe};
\draw[arr] (repeat2.east)   -- (check2.west)
           node[midway, above, font=\scriptsize] {yes};
\draw[arr] (repeat2.north)  -- ++(0,1.4) -| (judge2.north)
           node[midway, above, font=\scriptsize] {no};
\draw[arr] (check2.east)    -- (robust2.west)
           node[midway, above, font=\scriptsize] {yes};
\draw[arr] (check2.south)   -- (weak2.north)
           node[midway, right, font=\scriptsize] {no};

\end{tikzpicture}%
}

\caption{%
  \textbf{The two CAS frameworks.}
  \textbf{(a) CAS-gen:} a jailbreak candidate is accepted
  only after passing evaluation $\kgen$ consecutive times, filtering out
  chance successes.
  \textbf{(b) CAS-eval:} a fixed jailbreak prompt is
  re-evaluated $\keval$ independent times; it is classified as a \emph{consistent
  jailbreak} only if all $\keval$ verdicts are harmful, suppressing
  judge-stochasticity false positives.
  Setting $\kgen{=}\keval{=}1$ recovers standard single-shot ASR.
}
\label{fig:framework}
\end{figure}
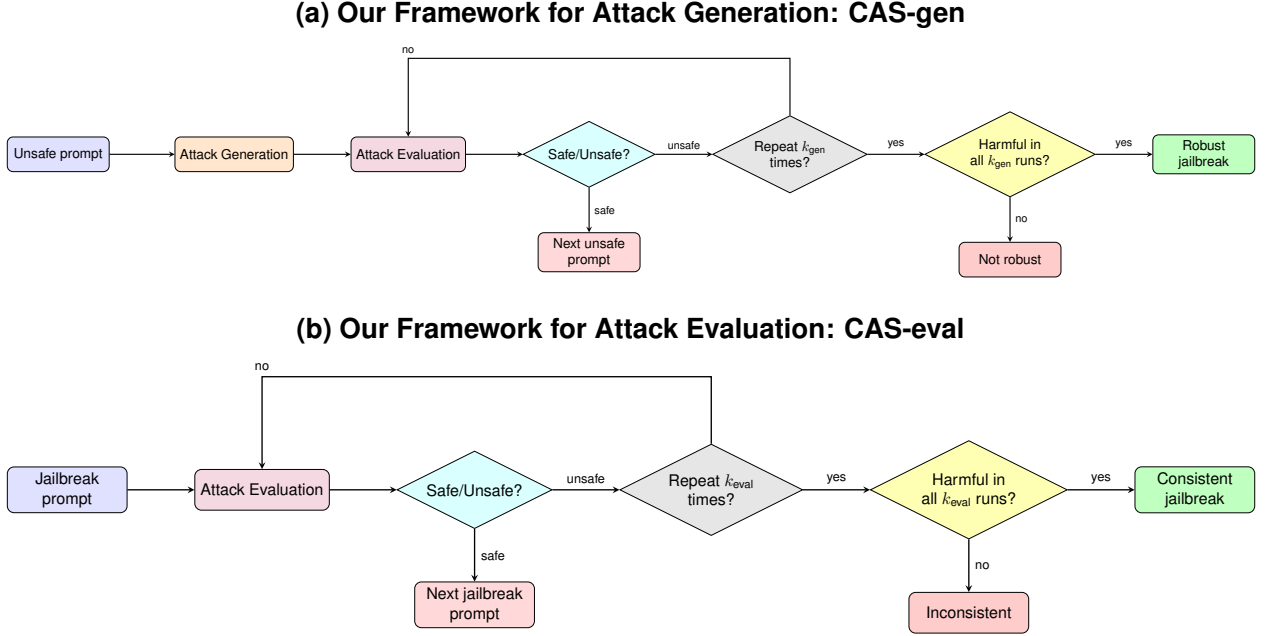

\subsubsection{A Novel Framework for Attack Evaluation (CAS-eval) (Figure~\ref{fig:framework}b)}
Given a fixed jailbreak prompt, the judge is queried $\keval$ independent times.
A prompt is classified as a \emph{consistent jailbreak} only if it passes all
$\keval$ evaluations, suppressing false positives from judge non-determinism.

In prior work, a jailbreak prompt is evaluated by the judge once ($\keval{=}1$):
a single harmful verdict is sufficient to count it as a successful attack.
Because the judge is stochastic at temperature $\theta > 0$, the same response
can receive different verdicts across repeated calls.
A single call may return ``harmful'' by chance even for a borderline response,
inflating the reported ASR without any genuine attack success.

CAS-eval addresses this by requiring $\cas(r, \keval) = 1$, classifying a prompt as
a consistent jailbreak only if all $\keval$ independent judge
calls return a harmful verdict.
This offers two concrete benefits over $\keval{=}1$.
First, it filters out false positives from judge non-determinism: a prompt that
passes all $\keval$ evaluations is genuinely harmful, whereas one that passes
only some is borderline and should not be counted as a success.
Second, it makes ASR estimates more reproducible across studies: because the
verdict no longer depends on a single stochastic draw, two independent
replications with the same $\keval$ will agree more closely.
Setting $\keval{=}1$ recovers the standard prior-work protocol exactly.

\section{Methodology for Evaluation}
\label{sec:methodology}

\subsection{Stochasticity Parameters}
\label{sec:notation}

Stochasticity can enter at either stage of attack generation or evaluation, but the parameters that govern it are
the same. Three parameters are central to our study.

\paragraph{Temperature} $T$ controls the diversity of the target model's outputs.
At $T{=}0$ the model is deterministic and will always\footnote{Our temperature analysis (see Appendix~\ref{app:temp-analysis}) showed that even under greedy decoding ($T=0$) some models exhibit non-determinism in generated output tokens.\label{note1}} produce the same response
to the same prompt; at $T{=}1$ outputs are maximally diverse, occasionally
sampling completions the model would suppress at lower temperatures.
As a result, ASR is sensitive to $T$. A jailbreak that succeeds 10\% of the
time at $T{=}0$ may succeed 40\% of the time at $T{=}1$, not because the attack
improved but because the model's sampling distribution widened.
Yet, most papers neither fix nor report $T$, making their results incomparable
across studies.
During the generation stage $T$ is denoted $\Tgen$; during evaluation, when
responses are re-generated to measure ASR sensitivity to temperature, it is
denoted $\Teval$.

\paragraph{Judge temperature} $\theta$ controls how stochastically the judge
assigns harmful/safe labels.
At $\theta{=}0$ the judge is fully\textsuperscript{\ref{note1}} deterministic: the same response always
receives the same verdict.
As $\theta$ increases, borderline responses (those whose content sits near
the harmful/safe boundary) become increasingly likely to flip between labels
across repeated calls.
Consequently, a measured ASR reflects not only whether responses are truly harmful,
but also which way the judge's random label draws happened to fall on the single
call recorded per response.
For example, suppose 10 responses are evaluated, 7 of which are clearly harmful
and always labeled as such, and 3 of which are borderline and labeled harmful
with probability 0.5.
In one run, the 3 borderline responses all draw ``harmful'', giving
$\mathrm{ASR} = 10/10 = 100\%$. In another run, they all draw ``safe'', giving
$\mathrm{ASR} = 7/10 = 70\%$, a 30-point swing from labeling noise alone,
with the attack and responses unchanged.
Two studies using the same judge model but different $\theta$ values are
effectively using different evaluation instruments, yet this parameter is rarely
reported.
During the generation stage $\theta$ governs the attack's internal candidate
selection ($\thetagen$); during the evaluation stage it governs post-hoc
labeling ($\theta_{\text{eval}}$).

\paragraph{Consistency threshold} $k$ is the number of consecutive harmful
verdicts the judge must return for a response to be counted as a successful
jailbreak.
The standard protocol uses $\keval{=}1$, where a single harmful verdict is sufficient.
Because both the target model and the judge are stochastic, a response that is
genuinely borderline will be counted as a success only some of the time:
re-generating the response at $\Teval > 0$ may produce a safer completion, and
even a fixed harmful response may be labeled safe when $\theta_{\text{eval}} > 0$.
Raising $k$ filters out this noise by requiring a response to clear the bar on every one
of $k$ consecutive calls, so only responses the judge \emph{consistently} labels
harmful are counted.
During the generation stage $k$ controls how stringently the attack's internal
filter selects candidates ($\kgen$); during the evaluation stage it controls how
stringently a response must pass the judge to be counted as a success ($\keval$).

\paragraph{Random seed} $\xi$ controls the stochastic sampling at both stages.
At temperature $T > 0$, the target model produces a different response to the
same prompt for each seed; at $\theta > 0$, the judge returns a different
verdict on the same response for each seed.
Most papers run a single seed and report the resulting ASR as if it were
deterministic, ignoring the seed-induced variance entirely.
In this work, we run $S$ independent seeds at the generation stage to quantify
how much ASR fluctuates across attack runs (generation stochasticity), and we
run $k$ independent judge calls to quantify how much the verdict fluctuates for
a fixed response (evaluation stochasticity).
The seed is therefore the mechanism through which both $T$ and $\theta$
translate into observable ASR variance.

\subsection{Metric: Attack Success Rate}

While \emph{CAS} gives a binary verdict per prompt, a paper ultimately reports
a single number summarizing attack performance across all $N$ evaluated prompts.
We define the \emph{Attack Success Rate} as the fraction of prompts for
which \emph{CAS} equals 1:
\begin{equation}
    ASR(\keval) = \frac{\sum_{n=1}^{N} \cas(r_n,\keval)}{N}
    \label{eq:asr_dataset}
\end{equation}

Equation~\ref{eq:asr_dataset} is the standard ASR formula, but parameterized by $\keval$.
The parameterization is the key contribution. As $\keval$ increases from 1 to 10,
$\text{ASR}(\keval)$ can drop sharply, not because the attack became less
effective, but because the stricter threshold filters out prompts that passed
under $\keval{=}1$ only by chance.
For instance, two papers reporting $\text{ASR}(\keval{=}1)$ and $\text{ASR}(\keval{=}10)$ on
the same attack can therefore show very different numbers, making the choice of
$\keval$ a direct confound for any cross-study comparison.

\subsection{Setup}
\textbf{Attacks.} We evaluate four black-box attacks:
Best-of-N~\cite{hughes2024bon} (repeated sampling),
PAIR~\cite{chao2023pair} (LLM-based red-teaming),
TAP~\cite{mehrotra2024tap} (tree-search LLM-based red-teaming), and
Crescendo~\cite{russinovich2024crescendo} (multi-turn LLM-based red-teaming).

\textbf{Target models.} We evaluate five instruction-tuned models spanning three
model families and three scales: Llama-3.2-1B-Instruct and Llama-3.1-8B-Instruct
and Llama-3.1-70B-Instruct (Meta), Gemma-3-1B-IT (Google), and
Granite-4.0-1B (IBM).

\textbf{Judges.} We use two Llama Guard models:
Llama-Guard-3-1B (LG-1B) and Llama-Guard-3-8B (LG-8B).

\textbf{Dataset.} All experiments use a subset of
JailbreakBench~\cite{chao2024jailbreakbench}, with $S{=}5$ independent
attack seeds (seeds 0--4) to quantify generation variance.

\section{Evaluation Stochasticity}
\label{sec:evaluation}

We evaluate four parameters at the attack evaluation stage ($\keval$, $\thetaeval$, $\Teval$, attack seed) and three at the attack generation stage ($\kgen$, $\Tgen$, $\thetagen$), and identify which parameters drive the largest variation in the reported ASR.

\subsection{Impact of Stochasticity at Attack Evaluation}
\label{sec:generation}

\begin{figure}[!htbp]
    \centering
    \includegraphics[width=\linewidth]{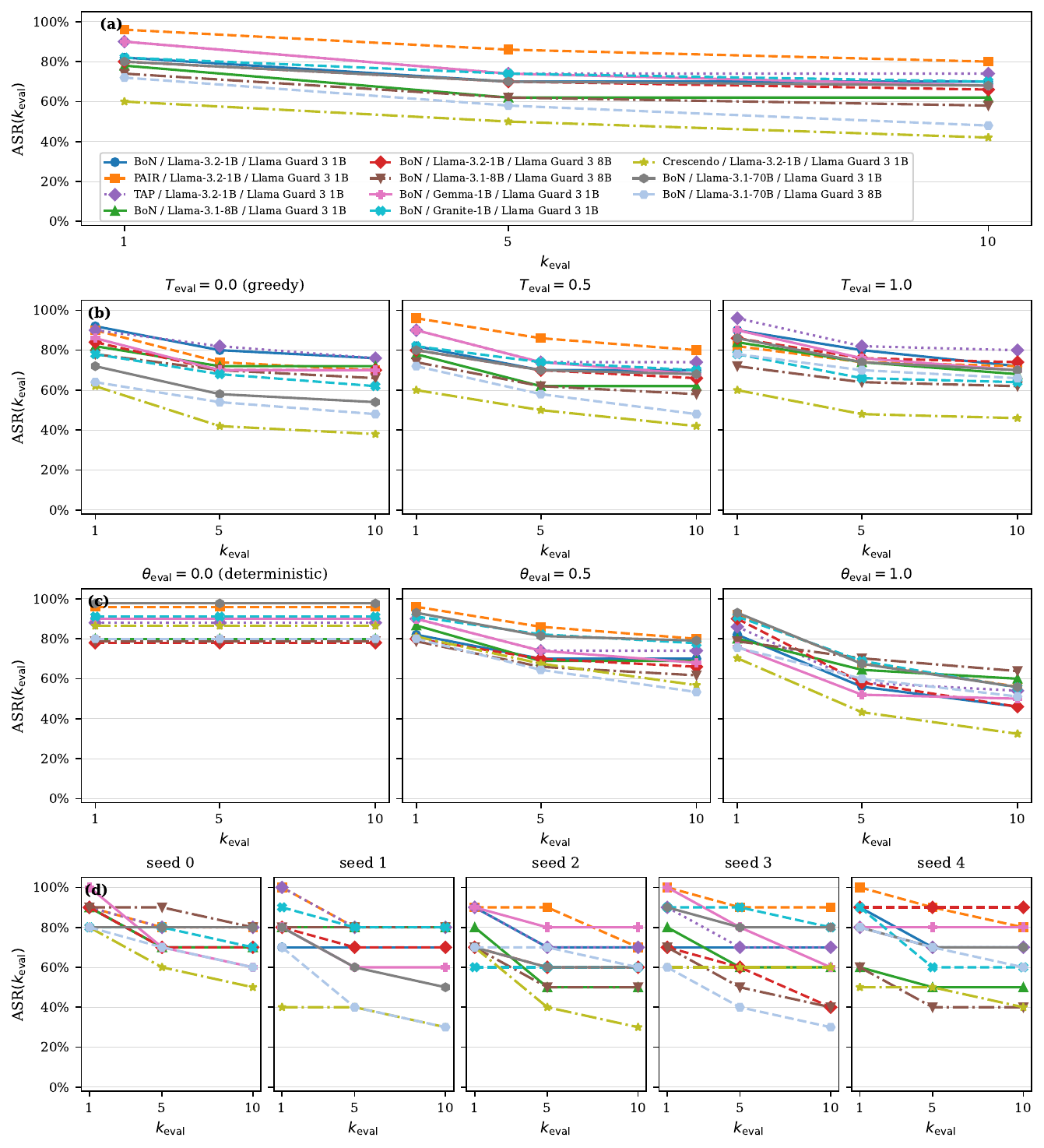}
    \caption{Impact of stochasticity parameters during attack evaluation on $\text{ASR}(\keval)$.
    Row (a): different models, judges, and attacks under the canonical condition.
    Row (b): influence of the target model temperature ($\Teval$).
    Row (c): influence of the judge temperature ($\thetaeval$).
    Row (d): impact of attack seed.}
    \label{fig:abl1_abl3}
\end{figure}

Figure~\ref{fig:abl1_abl3} reveals a clear hierarchy among the four parameters studied: evaluation threshold $\keval$, target model temperature $\Teval$, judge temperature $\thetaeval$, and attack seed.

% ---- Figure: Ablations 2 (canonical), 1, 3, and seed variance (4-row grid) ----

\paragraph{Effect of evaluation threshold $\keval$ (row~a)}
Moving from $\keval{=}1$ to $\keval{=}10$ reduces ASR by 12 to 24 percentage points across all 11 configurations (largest drop: BoN/Llama-3.1-70B/LG-8B at $-$24 pp; PAIR falls from 96\% to 80\%), so any absolute ASR figure reported under $\keval{=}1$ is upward-biased and cross-paper comparisons that mix $\keval$ values are invalid.
The relative ordering of attacks is preserved (PAIR $>$ TAP $>$ BoN), so qualitative conclusions survive the correction.
Judge choice further compounds the inflation: the LG-8B judge imposes an additional penalty that grows with target model scale, reaching 20 percentage points for the 70B target.
For unbiased absolute ASR estimates, $\keval{>}5$ should be used and reported. Qualitative attack rankings are safe regardless of $\keval$.

\paragraph{Effect of target model temperature $\Teval$ (row~b)}
Shifting $\Teval$ from 0.0 to 1.0 changes ASR by at most 18 percentage points with no systematic direction: different attacks peak at different temperatures, and no configuration reverses its relative rank across the three panels.
ASR comparisons are robust to the choice of $\Teval$, which need not be reported.

\paragraph{Effect of judge temperature $\thetaeval$ (row~c)}
At $\thetaeval{=}0$ every curve is perfectly flat. $\text{ASR}(\keval{=}10)$ equals $\text{ASR}(\keval{=}1)$ across all configurations, confirming that the monotonic decline in ASR with increasing $\keval$ is caused entirely by judge stochasticity.
Raising $\thetaeval$ to 1.0 collapses ASR by 15 to 54 percentage points across all 11 configurations; the largest drop is Crescendo at $\keval{=}10$, from 86\% ($\thetaeval{=}0$) to 32\% ($\thetaeval{=}1.0$), a reduction of 54 percentage points.
An unreported $\thetaeval$ is therefore a primary driver of inconsistency in published ASR numbers.

\paragraph{Effect of attack seed (row~d)}
Seed-to-seed spread is modest for most configurations (at most 20 percentage points), with the largest variation in LG-8B-judged BoN (50 percentage points at $\keval{=}10$), partly attributable to the small 10-prompt evaluation set where a single flip is worth 10 percentage points.
The relative ordering of configurations is preserved across seeds, and iterative refinement attacks (PAIR, TAP) are less sensitive to seed than single-pass sampling.
The attack seed need not be reported for qualitative comparisons; for absolute ASR estimates, averaging over at least 3 seeds is advisable to reduce the variance observed here.

\subsection{Impact of Stochasticity at Attack Generation}
\label{sec:generation2}

% ---- Figure: T_gen and theta_gen grid (3×3) ----
\begin{figure}[!htbp]
    \centering
    \includegraphics[width=\linewidth]{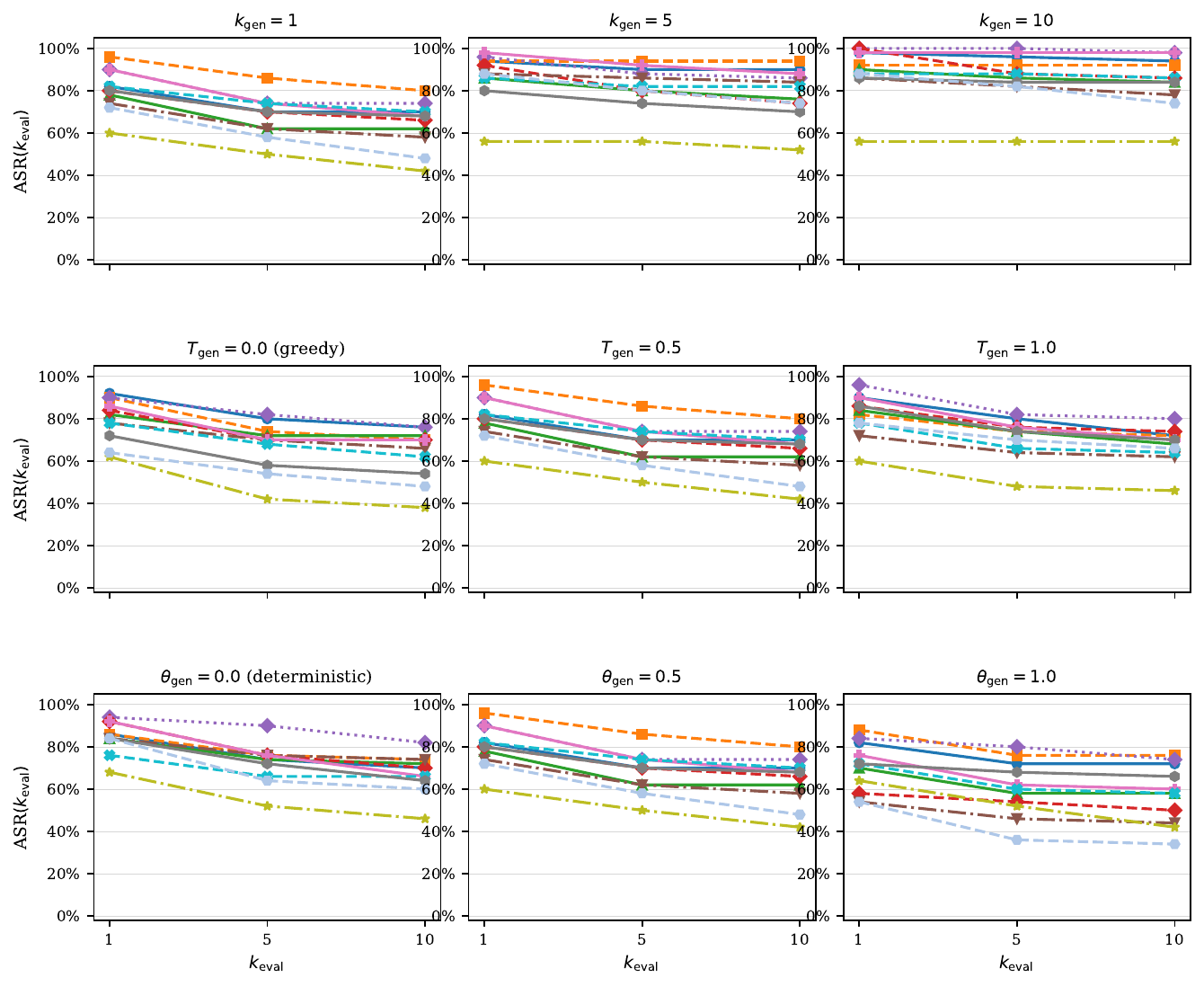}
    \caption{Effect of generation budget ($\kgen$), target model temperature ($\Tgen$),
    and judge generation temperature ($\thetagen$) on ASR$(\keval)$.
    Each column fixes one parameter value; curves show the four attack/judge configurations.}
    \label{fig:tgen_thetagen}
\end{figure}

Figure~\ref{fig:tgen_thetagen} reveals a clear hierarchy among the three parameters studied: generation budget $\kgen$, target model temperature $\Tgen$, and judge search temperature $\thetagen$.

\paragraph{Effect of generation budget $\kgen$ (row~a)}
Raising $\kgen$ from 1 to 10 lifts $\text{ASR}(\keval{=}10)$ by 12 to 30 percentage points across all configurations.
At $\kgen{=}10$, most LG-1B and PAIR/TAP curves become nearly flat, while LG-8B judged BoN variants retain a 12 to 14 pp decay; at $\kgen{=}1$, all curves decay by 12 to 24 pp from $\keval{=}1$ to $\keval{=}10$.
A tighter consistency filter selects unambiguously harmful responses that survive any post-hoc scrutiny; a loose filter ($\kgen{=}1$) accepts borderline responses that degrade sharply under repeated evaluation.
Two papers both claiming ``Best-of-N ASR'' can therefore differ by up to 30 pp at $\keval{=}10$ purely because of an undisclosed $\kgen$.

\paragraph{Effect of target model temperature $\Tgen$ (row~b)}
Shifting $\Tgen$ from 0.0 to 1.0 produces no systematic effect for 1B and 8B target models, with spreads of at most 14 pp.
Large (70B) target models show a mild monotonic increase, reaching at most 18 pp at $\keval{=}10$.
Target model temperature is therefore not a primary confound; it may be worth reporting only for very large target models.

\paragraph{Effect of judge search temperature $\thetagen$ (row~c)}
For PAIR, TAP, and Crescendo, and for BoN judged by LG-1B, the effect of $\thetagen$ is small (at most 8 pp spread at $\keval{=}10$).
BoN judged by LG-8B is the exception, with spreads up to 30 pp: a noisy search judge admits lower-quality responses that then fail post-hoc evaluation, and this noise compounds when the search and evaluation judge are the same model.
$\thetagen$ therefore only needs to be controlled for BoN with LG-8B; other attacks are unaffected by design.

\subsection{Takeaways}
\label{sec:takeaways}

Across all attacks, models, and judges, the experiments reveal a compact set of parameters that are worth standardizing and reporting.

\paragraph{When generating an attack.}
$\kgen$ is the single most consequential generation-side parameter: raising it from 1 to 10 lifts reported ASR by up to 30 percentage points, and two papers quoting ``Best-of-N ASR'' without disclosing $\kgen$ are not directly comparable.
$\thetagen$ matters only for BoN evaluated by LG-8B (up to 30 pp spread); other attacks are unaffected by design.
$\Tgen$ is a null result except for very large (70B+) target models, where a mild monotonic effect reaches at most 18 pp.

\paragraph{When evaluating an attack.}
$\thetaeval$ and $\keval$ jointly determine whether a reported ASR reflects genuine harmfulness or a lucky draw.
Setting $\thetaeval{=}0$ eliminates all judge stochasticity and makes $\text{ASR}(\keval{=}10)$ equal to $\text{ASR}(\keval{=}1)$; raising it to 1.0 can collapse ASR by up to 54 percentage points.
Using $\keval{>}1$ corrects the systematic upward bias of single-shot evaluation, which inflates ASR by 12 to 24 percentage points across all tested configurations.
$\Teval$ and attack seed contribute at most 18 and 20 percentage points respectively, with no systematic direction, and do not affect the relative ranking of attacks.

\paragraph{Minimum reporting checklist.}
At minimum, any paper reporting ASR should disclose $\kgen$, $\keval$, and $\thetaeval$.
For BoN attacks, $\thetagen$ is additionally required.
$\Tgen$, $\Teval$, and attack seed are secondary: reporting them is good practice but their omission rarely invalidates qualitative comparisons.

\section{Limitations}
\label{sec:limitations}

\paragraph{Attack coverage.}
Our experiments focus on Best-of-N (text), PAIR, TAP, and Crescendo.
Gradient-based attacks such as GCG operate differently, optimizing a
discrete suffix rather than sampling from the target model. Therefore, their
stochasticity profile may differ from the sampling-based attacks studied here.
Extending the analysis to GCG and to multimodal attacks (visual adversarial
examples, adversarial audio) is left for future work.

\paragraph{Target model coverage.}
All conclusions are drawn from experiments on Llama-3.2-1B, Llama-3.1-8B,
Llama-3.1-70B, Gemma3-1B, and Granite-3.2-1B.
These are all open-weight models with relatively limited safety alignment.
Whether the stochasticity effects reported here generalize to heavily RLHF-aligned proprietary models (e.g., GPT-4o, Claude) is an open
empirical question.
Beyond the model weights themselves, proprietary systems typically wrap the
model in a safety stack composed of rule-based filters, LLM classifiers, and
content moderation layers whose design is not publicly disclosed.
This opacity makes it difficult to isolate and measure stochasticity at the
generation and evaluation stages as our framework requires.
Furthermore, systematic jailbreak evaluation on proprietary models would likely
violate the terms of service of the corresponding API providers, so we
deliberately restrict our study to open-weight models.

\paragraph{Judge coverage.}
We evaluate LLM's responses with Llama-Guard-3-1B and Llama-Guard-3-8B, which are classifier-style judges. But, generative judges (such as the one proposed by StrongREJECT) may exhibit different temperature sensitivity. GPT-5.2-as-judge raises the same concern as proprietary target models: systematic
jailbreak evaluation through a commercial API would likely violate the terms of
service of the provider, so we deliberately exclude it from our study.

\paragraph{Scope of the proposed framework.}
CAS and the reporting standard address stochasticity at the generation and
evaluation stages.
They do not account for other sources of irreproducibility in jailbreak
research, such as system-prompt variation, tokenization differences across
API versions, or hardware non-determinism at $\Tgen{=}0$.

\section{Related Work}
\label{sec:related}

\paragraph{Safety Evaluation Reliability} Fraser et al.~\cite{fraser2025finetuning} documented surprising variance in safety
benchmark results from LLM stochasticity and trivial procedural variations, in the
context of fine-tuning. They showed that despite the cost, multiple runs are necessary to estimate random variation and get more robust evaluations. Our work extends this to the attack generation setting.
%They studied (i) the effect of the stochastic decoding with a non-zero temperature on the repeatability of the safety measurements in base and finetuned models; (ii)The effect of generation temperature on the evaluation of base and fine-tuned models. For all models, they generate responses at both temperature = 0 and temperature = 0.7. For a subset of cases, they repeat the temperature = 0.7 experiments five times, to better understand the variance at non-zero (non-deterministic) temperatures.

\paragraph{LLM Judge Reliability}
The closest work to ours on evaluation stochasticity is Schwinn et al.~\cite{schwinn2026coinflip},
who studied the reliability of safety rating using LLM judges. They showed that LLM judges are substantially less reliable for adversarial safety evaluations than previously assumed, performing on average only slightly better than a random coin-flip. 
Hence, they proposed techniques to mitigate judge's reliability. %corrected ASR by scaling results by the judge’s precision.
%They propose ReliableBench and JudgeStressTest. 
Our work differs in that we (1) study judge stochasticity as a function of temperature and number of evaluations, (2) jointly study generation stochasticity, and (3) focus specifically on BoN, PAIR, TAP, and Crescendo.

%\paragraph{Reproducibility in Jailbreak Benchmarking}
%Fang et al.~\cite{fang2026foundry} (Jailbreak Foundry) reproduced 30 attacks with a mean deviation of $+0.26$pp from reported ASR, treating ASR as a point estimate. Our work shows that this point estimate conceals large variance.

\paragraph{Statistical Modeling of ASR}
Feng et al.~\cite{feng2026saber} highlighted that ASR@N provides a more faithful measure of operational risk than ASR@1. Therefore, to reduce the need of testing attacks N (large) times, the authors proposed the SABER framework. SABER estimates ASR under Best-of-N sampling using a Beta distribution, predicting ASR@N from small-budget measurements (e.g., ASR@1000 from 100 samples). We adapt the SABER framework to our \textit{CAS} in Appendix~\ref{app:and-aggregation}.%The proposed scaling law enables reliable extrapolation of ASR, but does not answer the questions at hand in our work.

%\paragraph{Reproducibility in Machine Learning}
%\todo{Brief discussion of the broader ML reproducibility literature (Pineau et al., Bouthillier et al., etc.) and how our work connects to it.}

\section{Conclusion}
\label{sec:conclusion}

ASR is not a fixed property of an attack: it is a stochastic quantity jointly determined by judge temperature, evaluation threshold, generation budget, and attack seed.
Across four attacks and five target models, $\thetaeval$ alone can shift reported ASR by up to 54 pp, $\keval$ by 12 to 24 pp, $\kgen$ by 12 to 30 pp, and $\thetagen$ by 20 to 30 pp for BoN with a large judge.
Most published results fix none of these, making cross-paper comparisons statistically meaningless; we address this by introducing \textbf{CAS-gen} and \textbf{CAS-eval}, built on the \emph{Consistency for Attack Success} metric, and a minimum reporting standard whose required additions ($\kgen$, $\keval$, $\thetaeval$, confidence intervals) impose negligible overhead. The proposed frameworks will support the development of the MLCommons AI Security benchmarks.

\clearpage
\newpage
\bibliographystyle{unsrtnat}
\bibliography{references}

\newpage
\appendix
\appendix

% ============================================================
\section{Experiment}
\subsection{Experimental Hardware}
For our experiments, we use a cluster of two Nvidia A100 GPUs (80GB RAM each). Most of our experiment can be done with a single A100. Experiments requiring two A100 GPUs were the experiments involving a cumulation of large size LLMs (70B target model with a 8B judge model).

\subsection{Models}
All models are loaded from the Hugging Face Hub~\cite{huggingface_hub} using
the \texttt{transformers} library~\cite{wolf2020transformers}.
Table~\ref{tab:hf_models} lists the Hugging Face model identifier for each
model used in this paper.

\begin{table}[h]
\centering
\caption{Hugging Face model identifiers.}
\label{tab:hf_models}
\small
\begin{tabular}{lll}
\toprule
\textbf{Role} & \textbf{Model} & \textbf{Hugging Face ID} \\
\midrule
\multirow{5}{*}{Target} & Llama-3.2-1B-Instruct  & \texttt{meta-llama/Llama-3.2-1B-Instruct} \\
                        & Llama-3.1-8B-Instruct  & \texttt{meta-llama/Llama-3.1-8B-Instruct} \\
                        & Llama-3.1-70B-Instruct & \texttt{meta-llama/Llama-3.1-70B-Instruct} \\
                        & Gemma-3-1B-IT          & \texttt{google/gemma-3-1b-it} \\
                        & Granite-4.0-1B         & \texttt{ibm-granite/granite-4.0-1b} \\
\addlinespace
\multirow{2}{*}{Judge}  & Llama-Guard-3-1B       & \texttt{meta-llama/Llama-Guard-3-1B} \\
                        & Llama-Guard-3-8B       & \texttt{meta-llama/Llama-Guard-3-8B} \\
\bottomrule
\end{tabular}
\end{table}

\subsection{Experimental Parameters}
\label{app:parameters}
% ============================================================

Table~\ref{tab:parameters} lists all parameters used in our experimental framework,
together with their mathematical symbols, values or ranges, and descriptions.

\begin{table}[h]
\centering
\caption{Complete list of experimental parameters.}
\label{tab:parameters}
\small
\begin{tabular}{llll}
\toprule
\textbf{Symbol} & \textbf{Name} & \textbf{Value / Range} & \textbf{Description} \\
\midrule
$N$ & Candidate budget & 10{,}000 & Max augmented candidates per Best-of-N run; \\
    &                  &          & search stops early on first $\kgen$-consistent success. \\
\addlinespace
$\kgen$ & Attack consistency threshold & $\{1,5,10\}$ & Consecutive harmful judge labels required to accept \\
    &                              &                  & a candidate during search. Higher $\kgen$ = stricter. \\
\addlinespace
$\Tgen$ & Target model temperature & $\{0.0,0.5,1.0\}$ & Sampling temperature of the target model. \\
    &                          &                            & $\Tgen{=}0$ is greedy (deterministic). \\
\addlinespace
$\keval$ & Eval consistency threshold & $\{1,5,10\}$ & Consecutive independent judge evaluations required \\
    &                            &                  & post-hoc. $\text{ASR}(\keval)$ = fraction passing all $\keval$ evals. \\
\addlinespace
$\thetagen$ & Judge search temperature & $\{0.0,0.5,1.0\}$ & Judge temperature during attack generation \\
\addlinespace
$\theta_{\text{eval}}$ & Judge eval temperature & $\{0.0,0.5,1.0\}$ & Judge temperature during post-hoc evaluation. \\
\addlinespace
$S$ & Attack seeds & $5$ (seeds $0$--$4$) & Independent runs per prompt per condition. \\
    &              &                      & Used to assess the attack generation variance \\
\bottomrule
\end{tabular}
\end{table}

\subsection{Confidence Intervals}
\label{app:ci}

All ASR estimates are reported with 95\% Wilson score confidence intervals~\cite{wilson1927}.
Here, $n$ is the total number of (prompt, seed) pairs evaluated and $s$ is the number of those
pairs for which $\cas(r, \keval) = 1$, i.e., the runs counted as
successful jailbreaks.
The observed ASR is $\hat{p} = s/n$ and the Wilson interval is:
\begin{equation}
    \left[
      \frac{\hat{p} + \frac{z^2}{2n} \;\mp\; \frac{z}{1 + z^2/n}
            \sqrt{\frac{\hat{p}(1-\hat{p})}{n} + \frac{z^2}{4n^2}}}
           {1 + z^2/n}
    \right]
    \label{eq:wilson}
\end{equation}
where $n = N_{\text{prompts}} \times S$ is the total number of (prompt, seed) pairs
($N_{\text{prompts}} = 1{,}000$ prompts and $S = 5$ seeds, giving $n = 5{,}000$), and $z = 1.96$ is the
97.5th percentile of the standard normal distribution $\mathcal{N}(0,1)$,
chosen so that the two-sided interval $[\hat{p} - \epsilon,\, \hat{p} + \epsilon]$
covers the true proportion with probability 0.95~\cite{abramowitz1964handbook}.

The Wilson interval is preferred over the normal approximation (Wald interval)
because it remains valid when $\hat{p}$ is near 0 or 1, which is common when
ASR is very low (near-perfect defense) or very high (near-perfect attack).

\section{Statistical estimation of ASR for large k}
\label{app:and-aggregation}

In \cite{feng2026saber}, authors developed a statistical framework to estimate adversarial risk with scaled attack budget. It focuses on the standard Best-of-$N$ attack success rate, which uses an OR aggregation rule: a harmful query is counted as successfully jailbroken if \emph{at least one} out of $N$ independently sampled attempts succeeds. This metric is appropriate for modeling an attacker who is satisfied with any successful jailbreak (OR-aggregation). However, in our settings, one may instead care about \emph{consistency} or \emph{robust vulnerability}: a query should count as successfully attacked only if \emph{all} attempts succeed (AND-aggregation). This appendix develops the corresponding statistical formulation to predict the ASR for large $k$ under our setting.

\paragraph{Setup.}
For each harmful query $q_i$, let $S_{i,j} \in \{0,1\}$ denote the success indicator of the $j$-th attempt, where $S_{i,j}=1$ indicates a successful jailbreak. We assume that each query $q_i$ has a latent per-attempt success probability $p_i \in (0,1)$ and that, conditional on $p_i$, repeated attempts are i.i.d. Bernoulli:
\begin{equation}
    S_{i,1},S_{i,2},\dots \mid p_i \overset{\mathrm{i.i.d.}}{\sim} \mathrm{Bernoulli}(p_i).
    \label{eq:and-beta-bernoulli}
\end{equation}

Let's define the AND-aggregated success indicator for query $q_i$ under $N$ attempts as
\begin{equation}
    A_i^{(k)} := 
    \prod_{j=1}^{k} S_{i,j}.
    \label{eq:and-def}
\end{equation}
Thus, $A_i^{(k)}=1$ if and only if \emph{all} $k$ attempts succeed. Conditional on $p_i$, independence immediately gives
\begin{equation}
    \Pr\!
    \left(A_i^{(k)}=1\mid p_i\right)
    = \Pr\!\left(S_{i,1}=\cdots=S_{i,k}=1 \mid p_i\right)
    = p_i^k.
    \label{eq:and-cond}
\end{equation}

\begin{figure}[!htbp]
    \centering
    \includegraphics[width=0.85\linewidth]{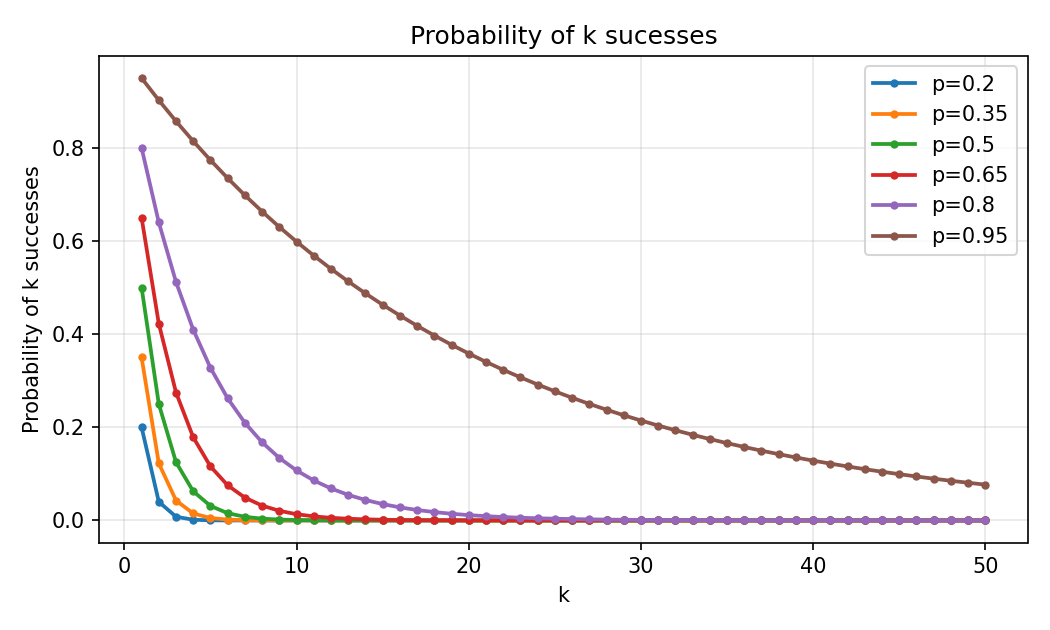}
    \caption{The trend of expected ASR as k scales.}
    \label{fig:stat}
\end{figure}

Assuming that the $p_i$ are themselves draws from some distribution $P$ over $[0,1]$, then for a dataset of $N$ prompts, the expected ASR will be 
$\mathbb{E}_{p\sim P}[p^k]$.

Figure~\ref{fig:stat} captures a notion of \emph{consistency robustness}. High ASR at large $k$ certifies reliable success across repeated attempts rather than occasional success. In the limit $k \to \infty$,
\[
\mathbb{E}_{p \sim P}[p^k] \to \Pr_{p \sim P}(p = 1),
\]
so the AND-aggregated ASR converges to the fraction of perfectly robust queries. Thus, the decay in ASR with increasing $k$ reflects instability in attack success and serves as a stress test for robustness under repeated adversarial pressure.

\clearpage

\section{Temperature Analysis}
\label{app:temp-analysis}
In this section, we present our analysis of the effect of temperature $T=0$ (greedy decoding) on the determinism of LLM output from the perspective of a commercial API. Indeed, it is commonly believed that an LLM with $T=0$ would consistently output the same answer when prompted with the same input.

\subsection{Dataset}
Since we could not use jailbreak prompts without breaching the terms of service of the API provider, we built our own non-adversarial dataset (Table~\ref{tab:determinism_prompts}). This dataset contains 20 prompts across five categories: factual, math, code, creative, and reasoning. Factual and math prompts have single correct answers and are expected to be highly deterministic. Creative prompts are open-ended and expected to show maximum variance.

\begin{table}[!h]
\centering
\caption{Prompt dataset used for the temperature-0 determinism study}
\label{tab:determinism_prompts}
\small
\begin{tabular}{lp{0.7\linewidth}}
\toprule
\textbf{Category} & \textbf{Prompt} \\
\midrule
Factual & What is the capital of France? \\
Factual & What is the chemical symbol for gold? \\
Factual & What are the three laws of thermodynamics? List them briefly. \\
Factual & What is the speed of light in meters per second? Give the exact value. \\
Factual & Name the four DNA nucleotide bases. \\
Factual & What were the main causes of World War I? Provide a concise summary. \\
\addlinespace
Math & What is 17 multiplied by 13? \\
Math & What is 2 raised to the power of 10? \\
Math & Calculate the area of a circle with radius 5. Use pi = 3.14159. \\
Math & What is the derivative of $f(x) = 3x^3 + 2x^2 - 5x + 7$? \\
\addlinespace
Code & Write a Python function \texttt{factorial} that computes $n!$ recursively. \\
Code & Write a one-line Python list comprehension for squares of 1 to 10. \\
Code & Write a Python function that returns a list of integers sorted descending without modifying the original. \\
Code & Write a Python function that checks whether a string is a palindrome (case-insensitive). \\
\addlinespace
Creative & Write a haiku about the ocean. \\
Creative & Give a product name and a one-sentence slogan for a fictional coffee brand. \\
Creative & Write the first sentence of a mystery novel set in 1920s Paris. \\
\addlinespace
Reasoning & If all mammals are warm-blooded and dolphins are mammals, what can we conclude? Explain. \\
Reasoning & A bat and a ball cost \$1.10. The bat costs \$1.00 more than the ball. How much is the ball? \\
Reasoning & Three mislabeled boxes (apples, oranges, both). One draw. How do you label all correctly? \\
\bottomrule
\end{tabular}
\end{table}

\subsection{Methodology}
\begin{figure}[h!]
    \centering
    \includegraphics[width=0.8\linewidth]{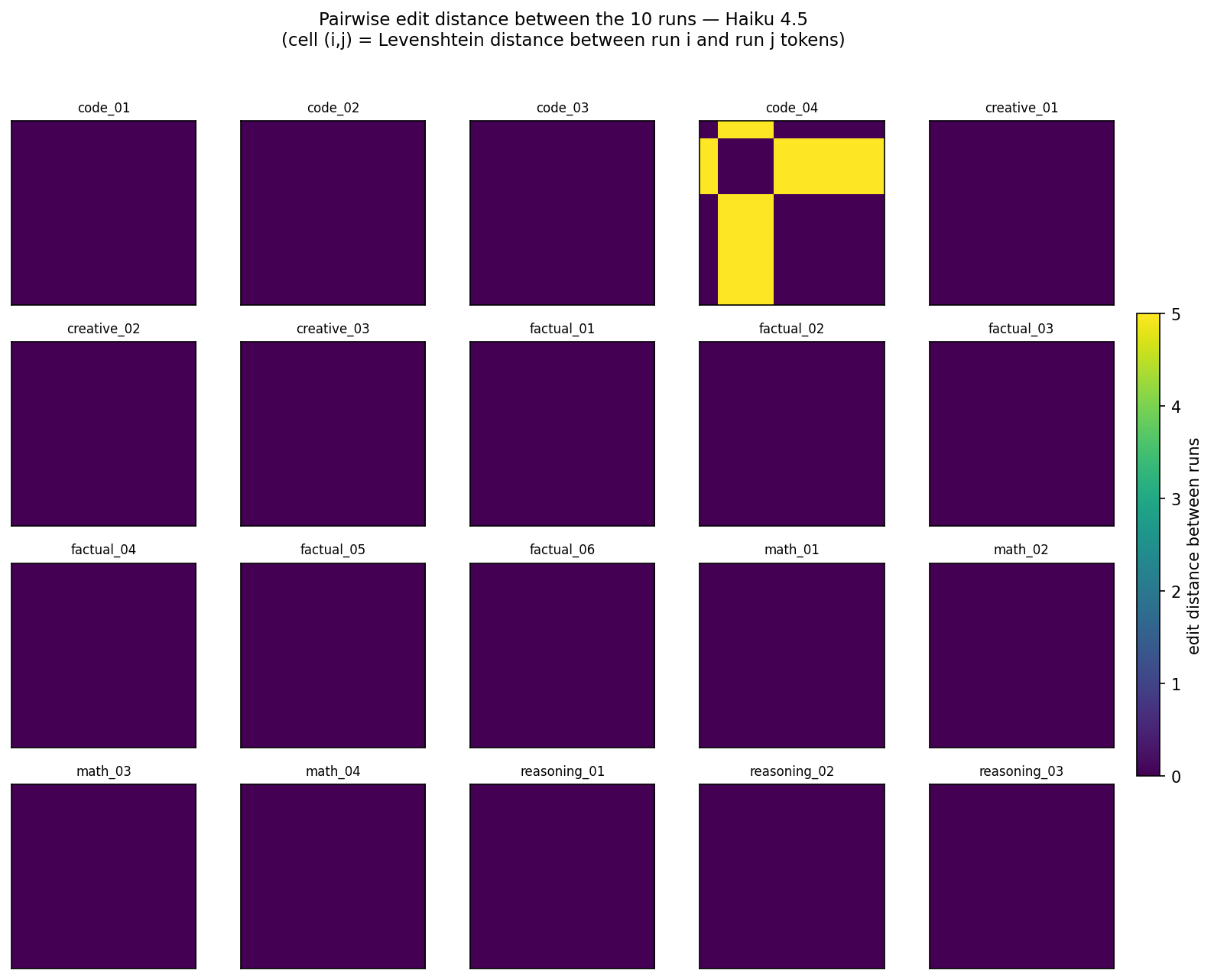}
    \caption{Temperature testing at $T=0$ for Haiku 4.5.}
    \label{fig:pairwise_distance_haiku}
\end{figure}

\begin{figure}[h!]
    \centering
    \includegraphics[width=0.8\linewidth]{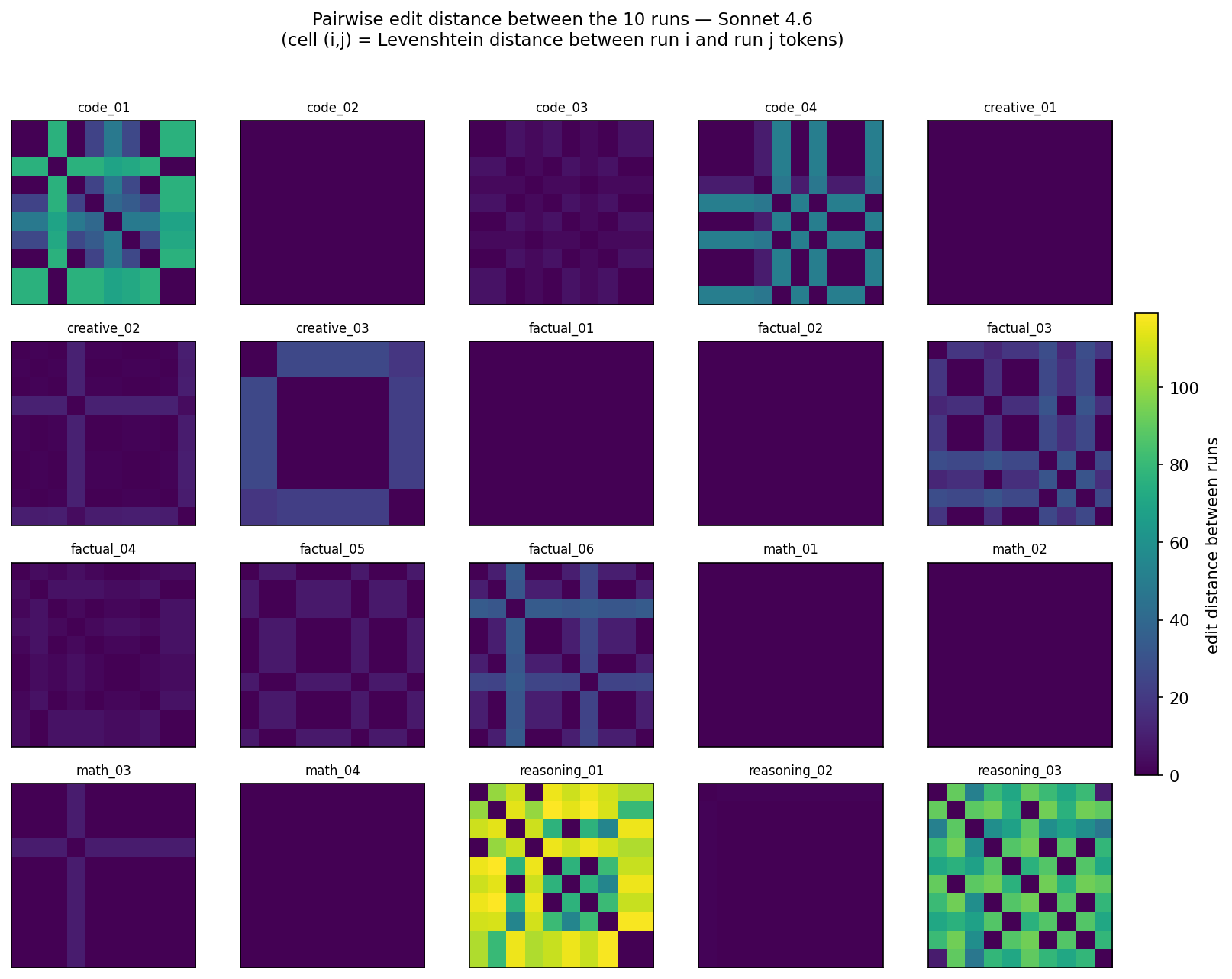}
    \caption{Temperature testing at $T=0$ for Sonnet 4.6.}
    \label{fig:pairwise_distance_sonnet}
\end{figure}

To assess the determinism of LLM output, we queried 10 times each prompt to selected models (Claude Sonnet 4.6 and Claude Haiku 4.5). From the responses, we measured the character-level edit distance between pairs (Levenshtein) as seen in Figure~\ref{fig:pairwise_distance_haiku} and Figure~\ref{fig:pairwise_distance_sonnet}. 

\subsection{Results}
We can make three observations based on Figure~\ref{fig:pairwise_distance_haiku} and Figure~\ref{fig:pairwise_distance_sonnet}. The first observation is temperature set to zero (T=0) does not guarantee determinism for commercial API. The second observation is the determinism depends on the prompt (see code\_04 in Figure~\ref{fig:pairwise_distance_haiku}). Lastly, determinism is model dependent for this commercial API. As a fact, Sonnet 4.6 exhibited significantly less determinism than Haiku 4.5, with 58.5\% vs 98.5\% exact-match rate.\\

\clearpage
% ============================================================
\section{Additional Results}
In this section, we provide plots specific to each target model and to each attacks studied in this paper. During our evaluation, we look at different angles.
In all bar charts and line plots, error bars and shaded bands show Wilson 95\% confidence intervals (see Appendix~\ref{app:ci}).

\paragraph{Different model sizes} we picked models from the same provider (Meta) and from the same generation (Llama 3). The goal was to see how the size of the target model (the model targeted by the jailbreak attack) had a stochastic impact during the generation of the attack and the evaluation of the attack. Also, we use different sizes for the model judge (Llama-Guard-3-1B and Llama-Guard-3-8B) to see how the settings of the judge can influence both attack generation and attack evaluation.

\paragraph{Different model providers} we picked models from the same provider (Meta) and from the same generation (Llama 3). The goal was to see how the size of the target model (the model targeted by the jailbreak attack) had a stochastic impact during the generation of the attack and the evaluation of the attack.

\paragraph{Different attacks} we picked models from the same provider (Meta) and from the same generation (Llama 3). The goal was to see how the size of the target model (the model targeted by the jailbreak attack) had a stochastic impact during the generation of the attack and the evaluation of the attack.
\clearpage
\subsection{Different model sizes}
\label{app:overview-model_sizes}

\subsubsection{ASR based consecutive judge evaluations}
\begin{figure}[h!]
  \centering
  \begin{tabular}{@{}m{0.02\linewidth}@{\hspace{4pt}}m{0.455\linewidth}@{\hspace{8pt}}m{0.455\linewidth}@{}}
    & \multicolumn{1}{c}{\small\textbf{Llama-Guard-3-1B}} & \multicolumn{1}{c}{\small\textbf{Llama-Guard-3-8B}} \\[4pt]
    \rotatebox[origin=c]{90}{\small\textbf{Llama-3.2-1B}} &
      \includegraphics[width=\linewidth,height=4.5cm,keepaspectratio]{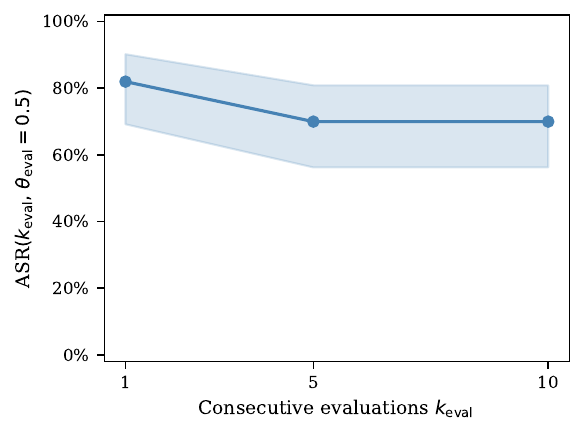} &
      \includegraphics[width=\linewidth,height=4.5cm,keepaspectratio]{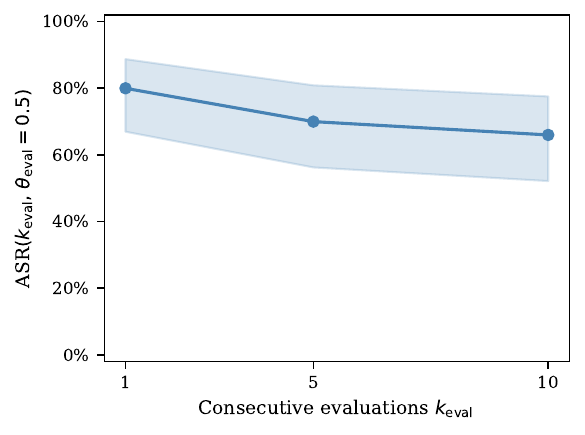} \\[4pt]
    \rotatebox[origin=c]{90}{\small\textbf{Llama-3.1-8B}} &
      \includegraphics[width=\linewidth,height=4.5cm,keepaspectratio]{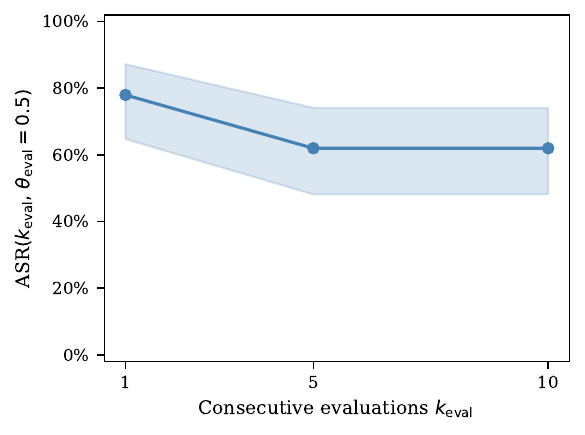} &
      \includegraphics[width=\linewidth,height=4.5cm,keepaspectratio]{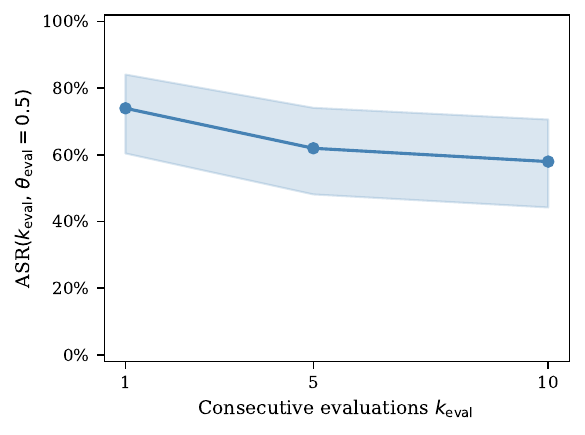} \\[4pt]
    \rotatebox[origin=c]{90}{\small\textbf{Llama-3.1-70B}} &
      \includegraphics[width=\linewidth,height=4.5cm,keepaspectratio]{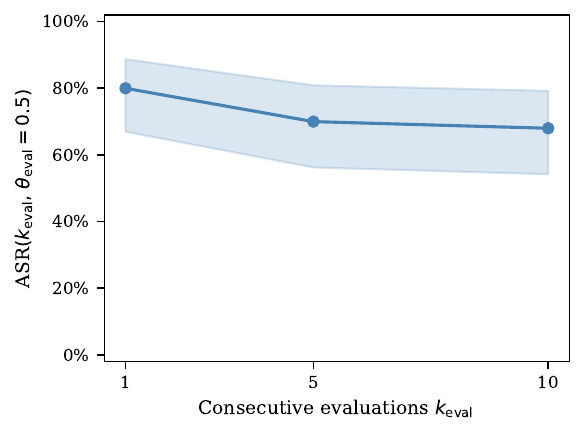} &
      \includegraphics[width=\linewidth,height=4.5cm,keepaspectratio]{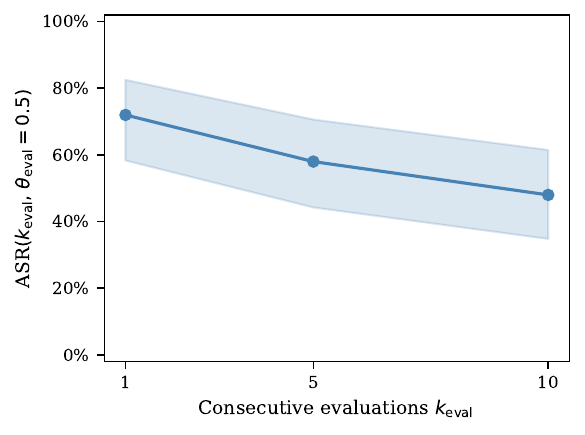} \\
  \end{tabular}
  \caption{\textbf{ASR($k_{eval}$)}; Parameters: $\kgen{=}1, \Tgen{=}0.5, \thetagen{=}0.5, \Teval{=}0.5, \theta_{\text{eval}}{=}0.5$}
  \label{fig:ov_abl2}
\end{figure}
\clearpage

\subsubsection{ASR based on target model temperature at evaluation}
\begin{figure}[h!]
  \centering
  \begin{tabular}{@{}m{0.02\linewidth}@{\hspace{4pt}}m{0.455\linewidth}@{\hspace{8pt}}m{0.455\linewidth}@{}}
    & \multicolumn{1}{c}{\small\textbf{Llama-Guard-3-1B}} & \multicolumn{1}{c}{\small\textbf{Llama-Guard-3-8B}} \\[4pt]
    \rotatebox[origin=c]{90}{\small\textbf{Llama-3.2-1B}} &
      \includegraphics[width=\linewidth,height=4.5cm,keepaspectratio]{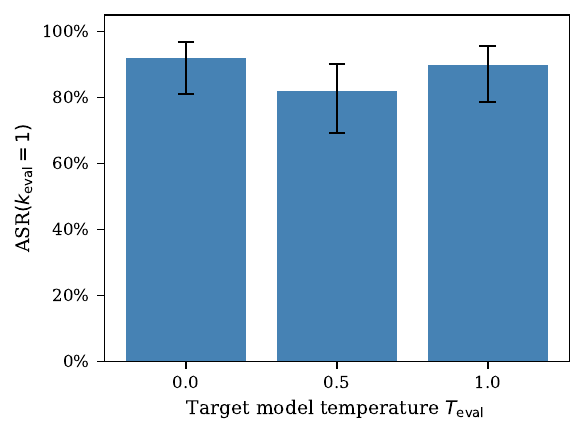} &
      \includegraphics[width=\linewidth,height=4.5cm,keepaspectratio]{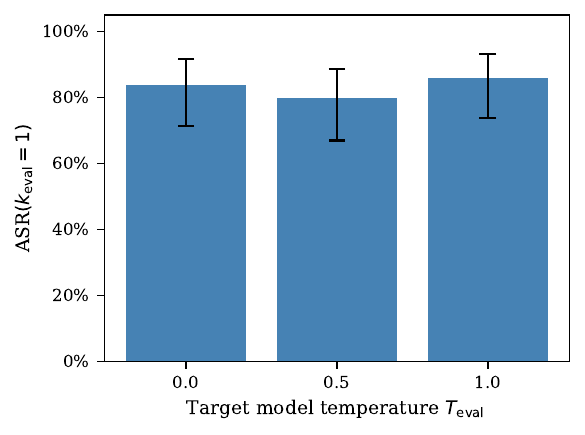} \\[4pt]
    \rotatebox[origin=c]{90}{\small\textbf{Llama-3.1-8B}} &
      \includegraphics[width=\linewidth,height=4.5cm,keepaspectratio]{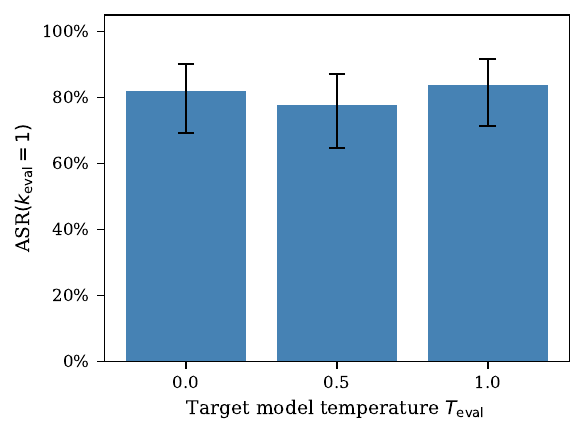} &
      \includegraphics[width=\linewidth,height=4.5cm,keepaspectratio]{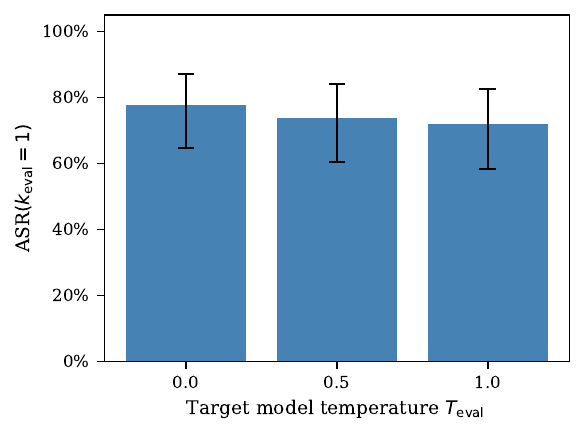} \\[4pt]
    \rotatebox[origin=c]{90}{\small\textbf{Llama-3.1-70B}} &
      \includegraphics[width=\linewidth,height=4.5cm,keepaspectratio]{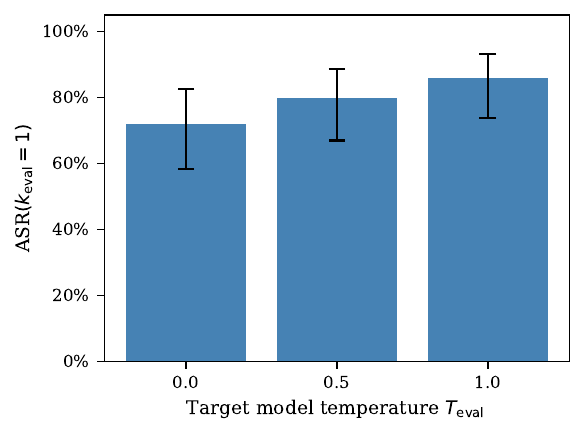} &
      \includegraphics[width=\linewidth,height=4.5cm,keepaspectratio]{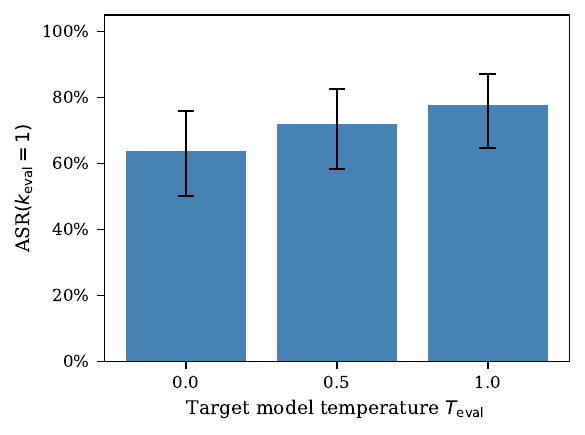} \\
  \end{tabular}
  \caption{\textbf{ASR($k_{eval}=1$, $T_{eval}$)}; Parameters: $\kgen{=}1, \Tgen{=}0.5, \thetagen{=}0.5,\theta_{\text{eval}}{=}0.5$}
  \label{fig:ov_abl1}
\end{figure}
\clearpage

\subsubsection{ASR based on judge model temperature at evaluation}

\begin{figure}[h!]
  \centering
  \begin{tabular}{@{}m{0.02\linewidth}@{\hspace{4pt}}m{0.455\linewidth}@{\hspace{8pt}}m{0.455\linewidth}@{}}
    & \multicolumn{1}{c}{\small\textbf{Llama-Guard-3-1B}} & \multicolumn{1}{c}{\small\textbf{Llama-Guard-3-8B}} \\[4pt]
    \rotatebox[origin=c]{90}{\small\textbf{Llama-3.2-1B}} &
      \includegraphics[width=\linewidth]{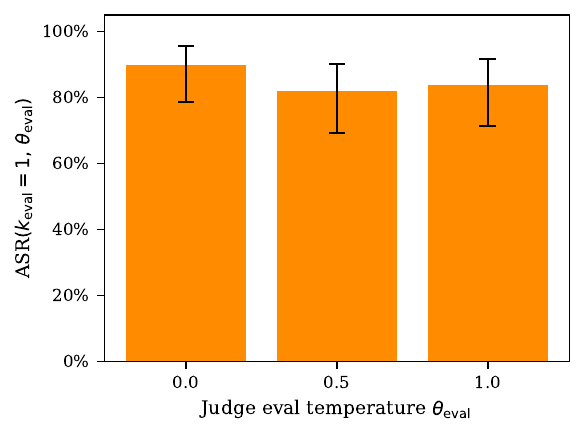} &
      \includegraphics[width=\linewidth]{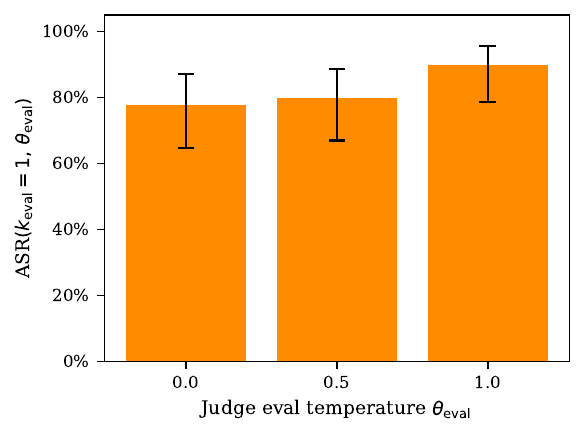} \\[4pt]
    \rotatebox[origin=c]{90}{\small\textbf{Llama-3.1-8B}} &
      \includegraphics[width=\linewidth]{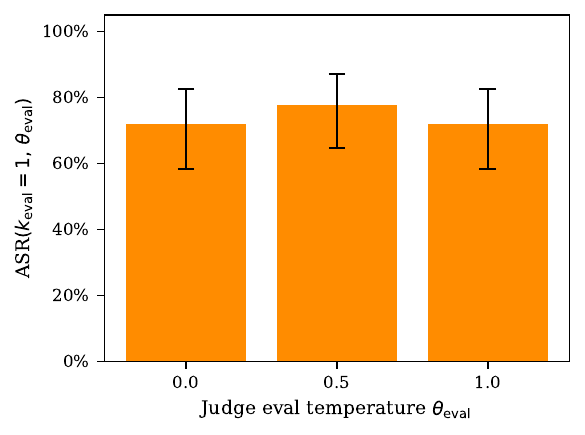} &
      \includegraphics[width=\linewidth]{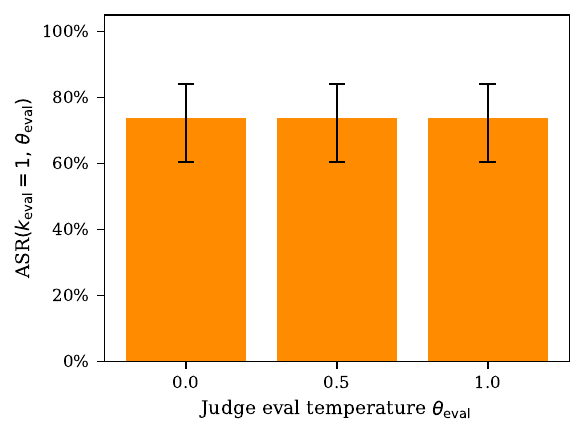} \\[4pt]
    \rotatebox[origin=c]{90}{\small\textbf{Llama-3.1-70B}} &
      \includegraphics[width=\linewidth]{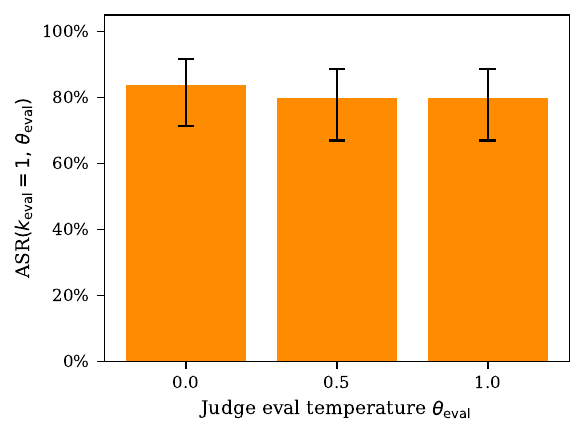} &
      \includegraphics[width=\linewidth]{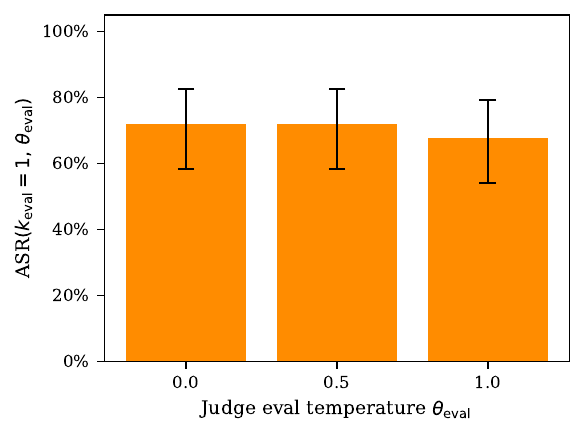} \\
  \end{tabular}
  \caption{\textbf{ASR($k_{eval}=1$, $\theta_{\text{eval}}$)}; Parameters: $\kgen{=}1, \Tgen{=}0.5, \thetagen{=}0.5, \Teval{=}0.5$}
  \label{fig:ov_abl3}
\end{figure}

\clearpage

\subsubsection{Heatmap ASR based on consecutive evaluations at generation}
\begin{figure}[h!]
  \centering
  \begin{tabular}{@{}m{0.02\linewidth}@{\hspace{4pt}}m{0.455\linewidth}@{\hspace{8pt}}m{0.455\linewidth}@{}}
    & \multicolumn{1}{c}{\small\textbf{Llama-Guard-3-1B}} & \multicolumn{1}{c}{\small\textbf{Llama-Guard-3-8B}} \\[4pt]
    \rotatebox[origin=c]{90}{\small\textbf{Llama-3.2-1B}} &
      \includegraphics[width=\linewidth,height=4.5cm,keepaspectratio]{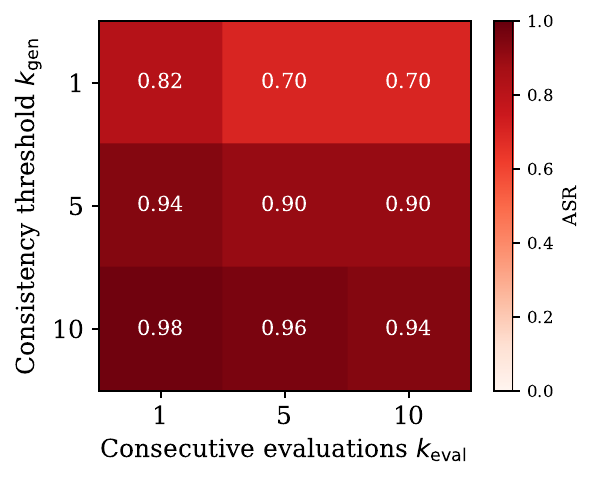} &
      \includegraphics[width=\linewidth,height=4.5cm,keepaspectratio]{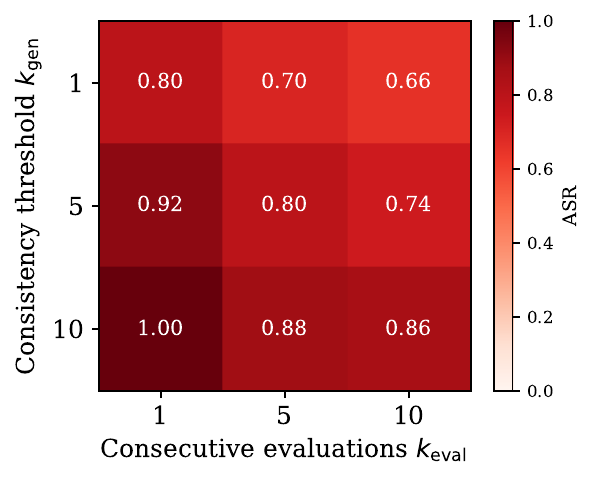} \\[4pt]
    \rotatebox[origin=c]{90}{\small\textbf{Llama-3.1-8B}} &
      \includegraphics[width=\linewidth,height=4.5cm,keepaspectratio]{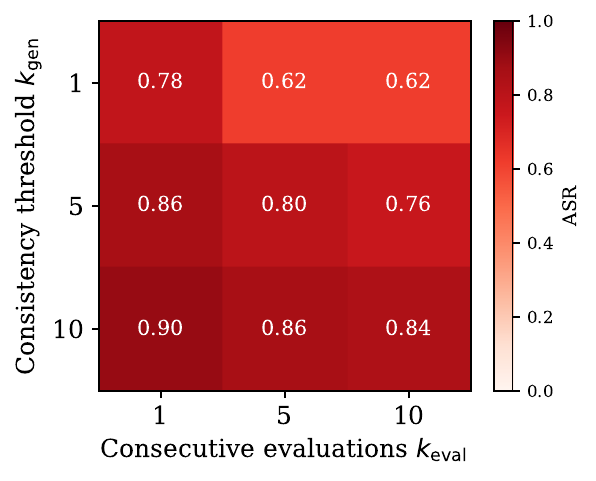} &
      \includegraphics[width=\linewidth,height=4.5cm,keepaspectratio]{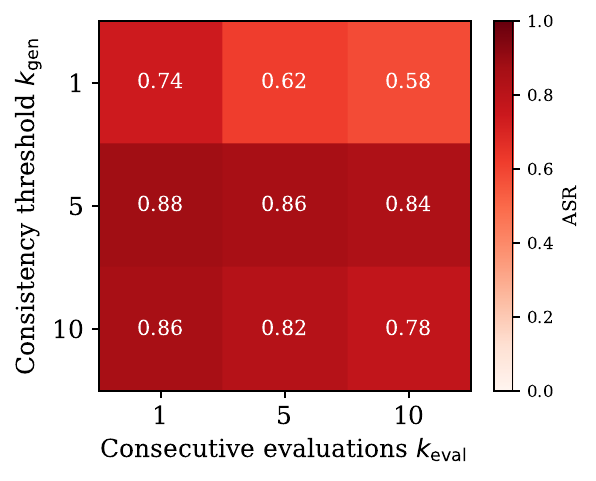} \\[4pt]
    \rotatebox[origin=c]{90}{\small\textbf{Llama-3.1-70B}} &
      \includegraphics[width=\linewidth,height=4.5cm,keepaspectratio]{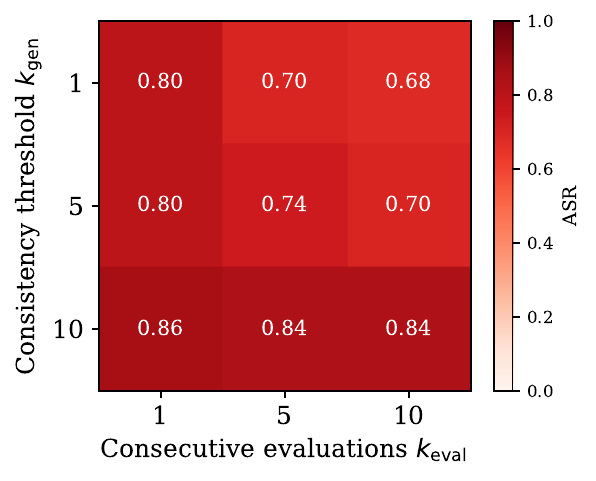} &
      \includegraphics[width=\linewidth,height=4.5cm,keepaspectratio]{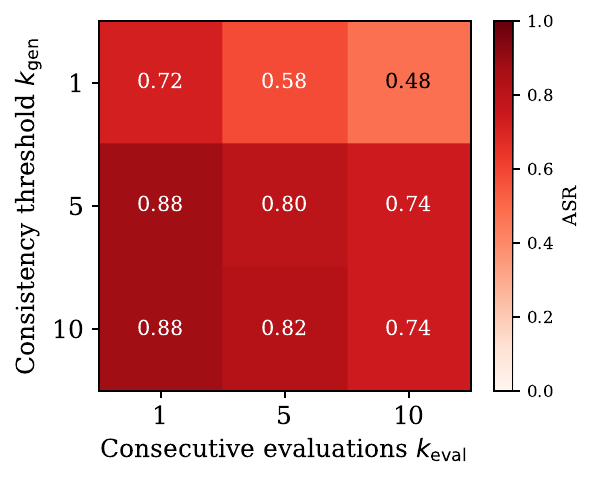} \\
  \end{tabular}
  \caption{Heatmap ASR($k_{eval},k_{gen}$) ; Parameters: $\Tgen{=}0.5, \thetagen{=}0.5, \Teval{=}0.5$}
  \label{fig:ov_abl5}
\end{figure}

\clearpage

\subsubsection{Heatmap ASR based on the temperature of the target model at generation}
\begin{figure}[h!]
  \centering
  \begin{tabular}{@{}m{0.02\linewidth}@{\hspace{4pt}}m{0.455\linewidth}@{\hspace{8pt}}m{0.455\linewidth}@{}}
    & \multicolumn{1}{c}{\small\textbf{Llama-Guard-3-1B}} & \multicolumn{1}{c}{\small\textbf{Llama-Guard-3-8B}} \\[4pt]
    \rotatebox[origin=c]{90}{\small\textbf{Llama-3.2-1B}} &
      \includegraphics[width=\linewidth]{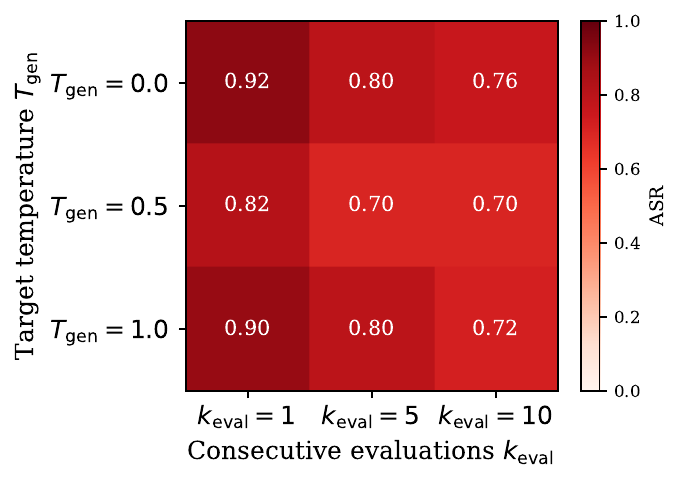} &
      \includegraphics[width=\linewidth]{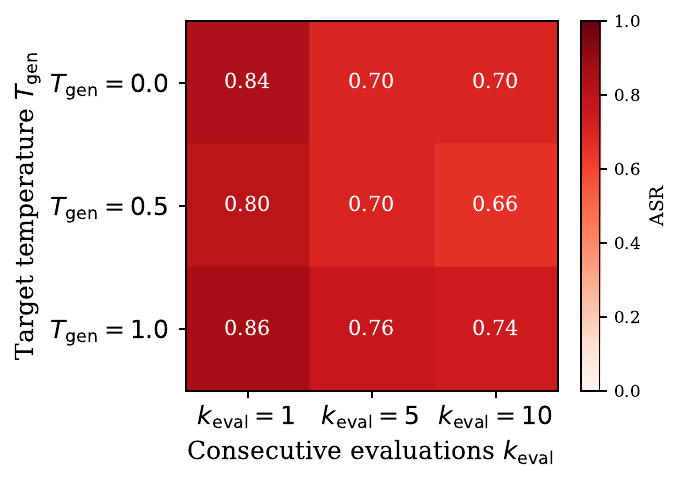} \\[4pt]
    \rotatebox[origin=c]{90}{\small\textbf{Llama-3.1-8B}} &
      \includegraphics[width=\linewidth]{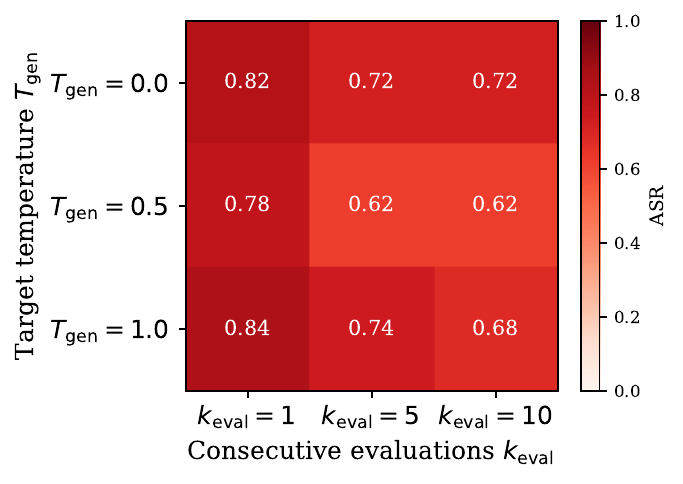} &
      \includegraphics[width=\linewidth]{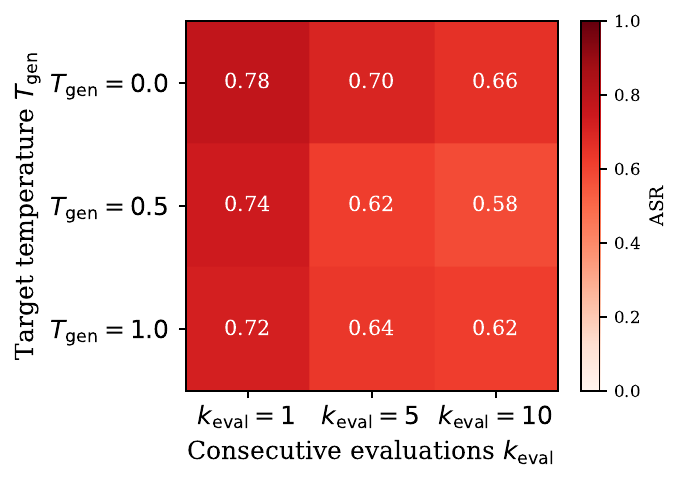} \\[4pt]
    \rotatebox[origin=c]{90}{\small\textbf{Llama-3.1-70B}} &
      \includegraphics[width=\linewidth]{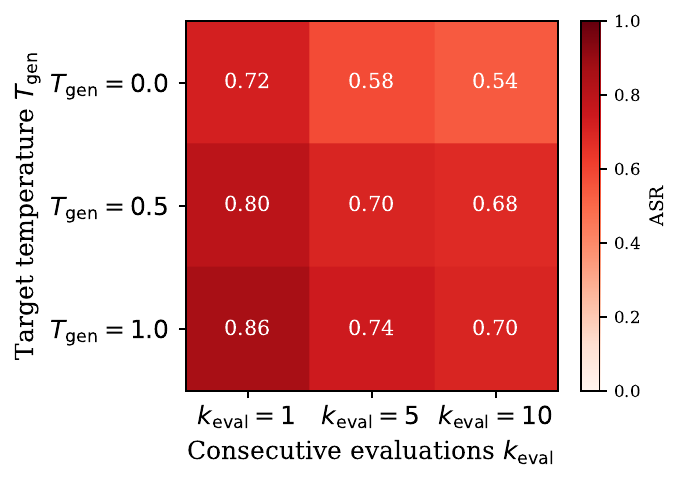} &
      \includegraphics[width=\linewidth]{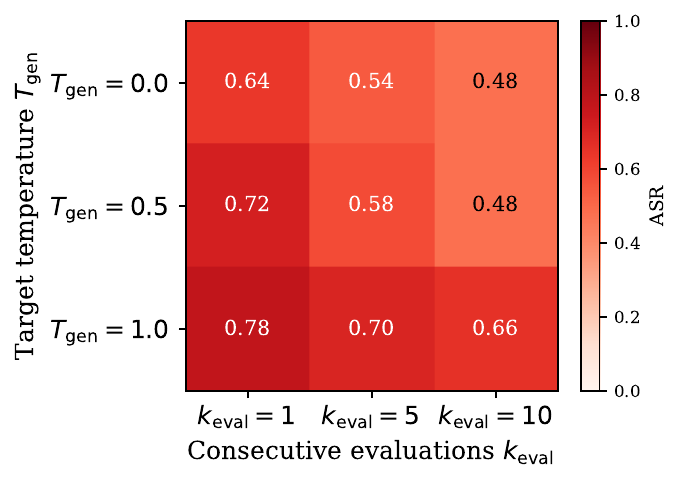} \\
  \end{tabular}
  \caption{Heatmap ASR($k_{eval},\Tgen$) ; Parameters: $\thetagen{=}0.5, k_{gen}=1, \Teval{=}0.5, \thetaeval{=}0.5$}
  \label{fig:ov_abl6}
\end{figure}

\clearpage
\subsubsection{Heatmap ASR based on the temperature of the judge at generation}
\begin{figure}[h!]
  \centering
  \begin{tabular}{@{}m{0.02\linewidth}@{\hspace{4pt}}m{0.455\linewidth}@{\hspace{8pt}}m{0.455\linewidth}@{}}
    & \multicolumn{1}{c}{\small\textbf{Llama-Guard-3-1B}} & \multicolumn{1}{c}{\small\textbf{Llama-Guard-3-8B}} \\[4pt]
    \rotatebox[origin=c]{90}{\small\textbf{Llama-3.2-1B}} &
      \includegraphics[width=\linewidth]{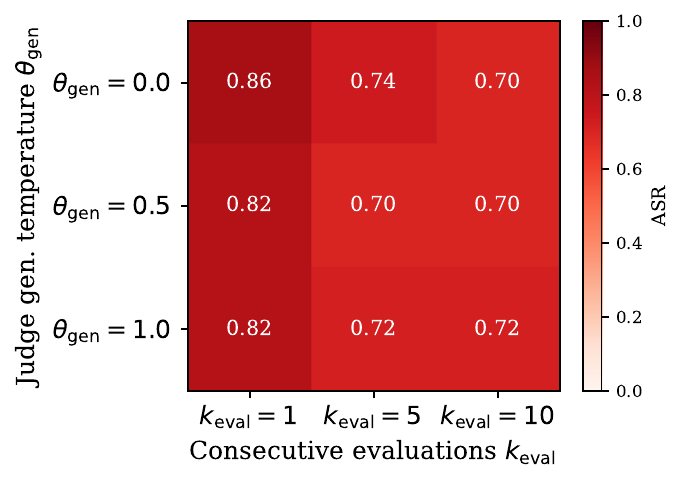} &
      \includegraphics[width=\linewidth]{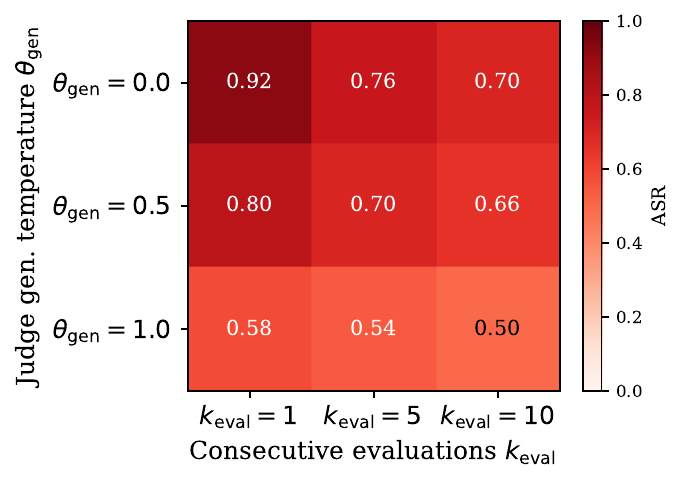} \\[4pt]
    \rotatebox[origin=c]{90}{\small\textbf{Llama-3.1-8B}} &
      \includegraphics[width=\linewidth]{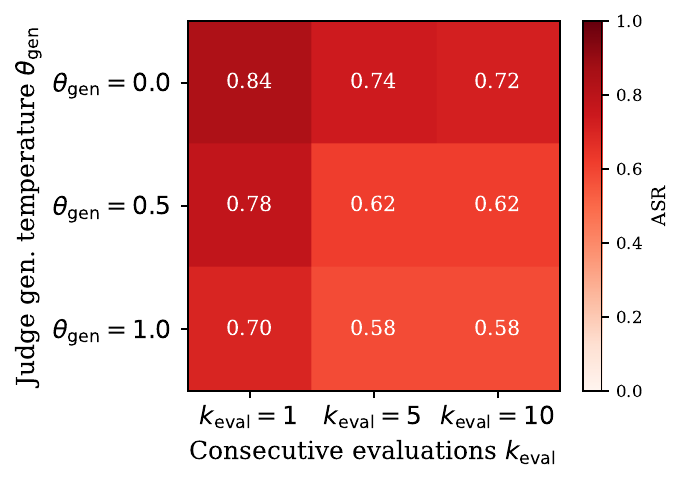} &
      \includegraphics[width=\linewidth]{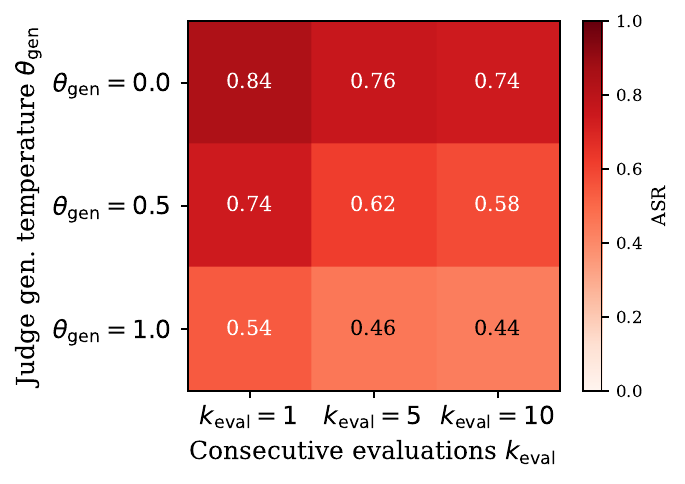} \\[4pt]
    \rotatebox[origin=c]{90}{\small\textbf{Llama-3.1-70B}} &
      \includegraphics[width=\linewidth]{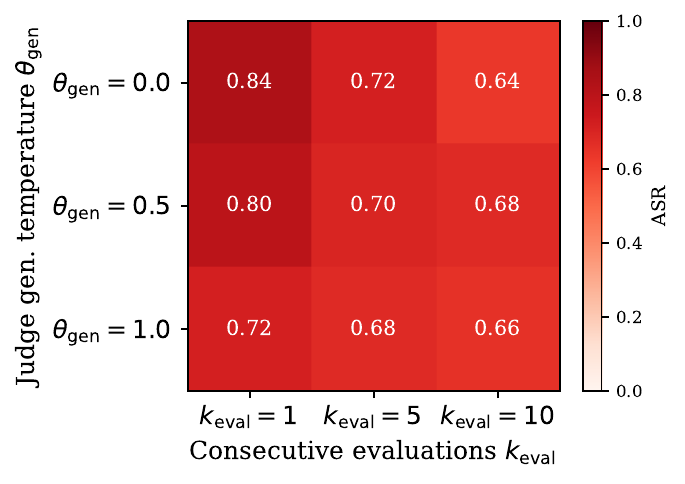} &
      \includegraphics[width=\linewidth]{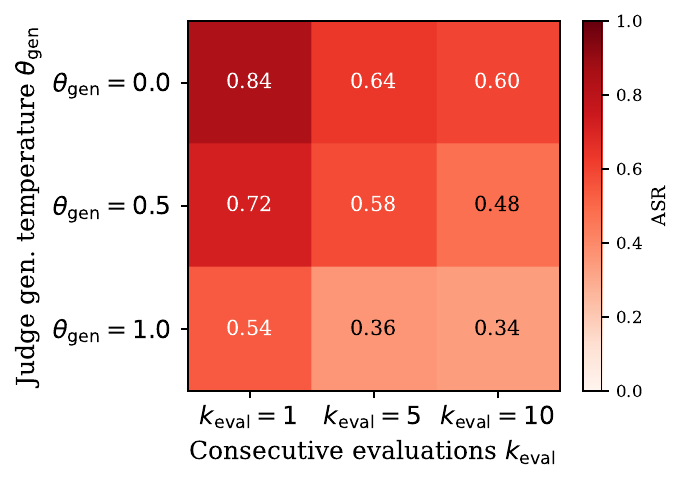} \\
  \end{tabular}
    \caption{Heatmap ASR($k_{eval},\thetagen$) ; Parameters: $\Tgen{=}0.5, k_{gen}=1, \Teval{=}0.5, \thetaeval{=}0.5$}
  \label{fig:ov_abl7}
\end{figure}

\clearpage

\subsection{Different model providers}
\label{app:overview-model_providers}

\subsubsection{ASR based consecutive judge evaluations}
\begin{figure}[h!]
  \centering
  \begin{tabular}{@{}m{0.3\linewidth}@{\hspace{8pt}}m{0.3\linewidth}@{\hspace{8pt}}m{0.3\linewidth}@{}}
    \multicolumn{1}{c}{\small\textbf{Llama-3.2-1B}} & \multicolumn{1}{c}{\small\textbf{Gemma3-1B}} & \multicolumn{1}{c}{\small\textbf{Granite-3.2-1B}} \\[4pt]
    \includegraphics[width=\linewidth]{assets/figures/abl2_asr_k.pdf} &
    \includegraphics[width=\linewidth]{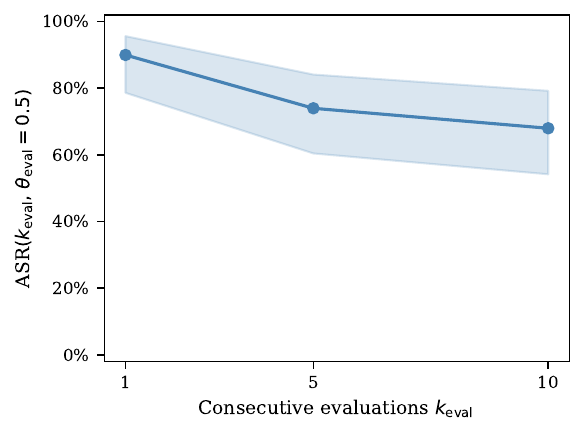} &
    \includegraphics[width=\linewidth]{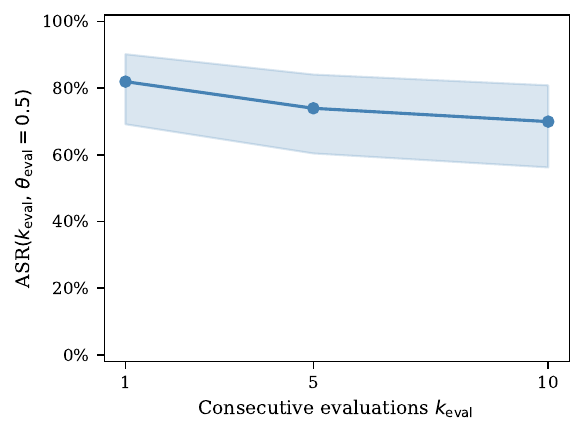} \\
  \end{tabular}
  \caption{\textbf{ASR($k_{eval}$)}; Parameters: $\kgen{=}1, \Tgen{=}0.5, \thetagen{=}0.5, \Teval{=}0.5, \theta_{\text{eval}}{=}0.5$}
  \label{fig:models_abl2}
\end{figure}

\subsubsection{ASR based on target model temperature at evaluation}
\begin{figure}[h!]
  \centering
  \begin{tabular}{@{}m{0.3\linewidth}@{\hspace{8pt}}m{0.3\linewidth}@{\hspace{8pt}}m{0.3\linewidth}@{}}
    \multicolumn{1}{c}{\small\textbf{Llama-3.2-1B}} & \multicolumn{1}{c}{\small\textbf{Gemma3-1B}} & \multicolumn{1}{c}{\small\textbf{Granite-3.2-1B}} \\[4pt]
    \includegraphics[width=\linewidth]{assets/figures/abl1_asr_T.pdf} &
    \includegraphics[width=\linewidth]{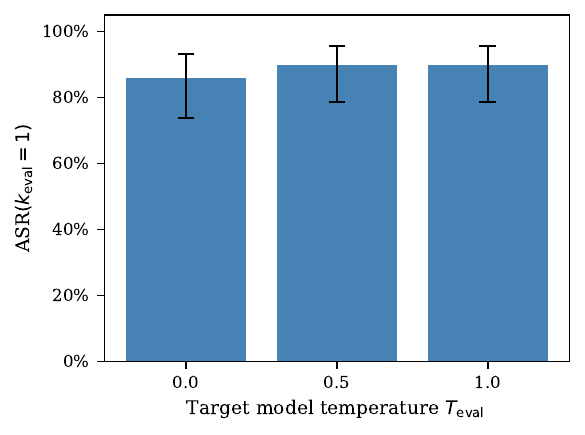} &
    \includegraphics[width=\linewidth]{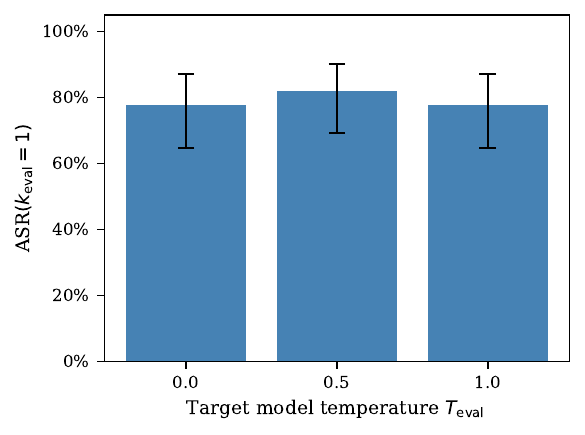} \\
  \end{tabular}
  \caption{\textbf{ASR($k_{eval}=1$, $T_{eval}$)}; Parameters: $\kgen{=}1, \Tgen{=}0.5, \thetagen{=}0.5,\theta_{\text{eval}}{=}0.5$}
  \label{fig:models_abl1}
\end{figure}

\subsubsection{ASR based on judge model temperature at evaluation}
\begin{figure}[h!]
  \centering
  \begin{tabular}{@{}m{0.3\linewidth}@{\hspace{8pt}}m{0.3\linewidth}@{\hspace{8pt}}m{0.3\linewidth}@{}}
    \multicolumn{1}{c}{\small\textbf{Llama-3.2-1B}} & \multicolumn{1}{c}{\small\textbf{Gemma3-1B}} & \multicolumn{1}{c}{\small\textbf{Granite-3.2-1B}} \\[4pt]
    \includegraphics[width=\linewidth]{assets/figures/abl3_asr_theta_eval.pdf} &
    \includegraphics[width=\linewidth]{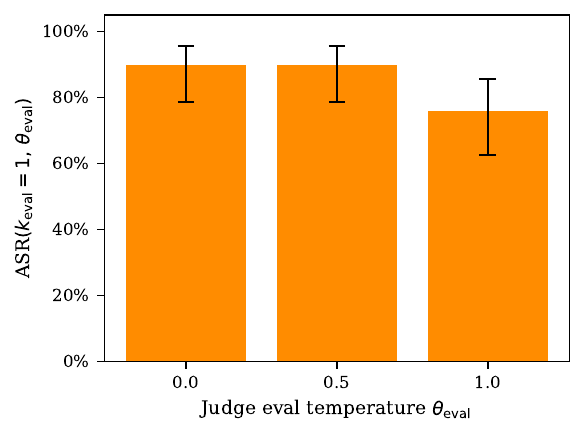} &
    \includegraphics[width=\linewidth]{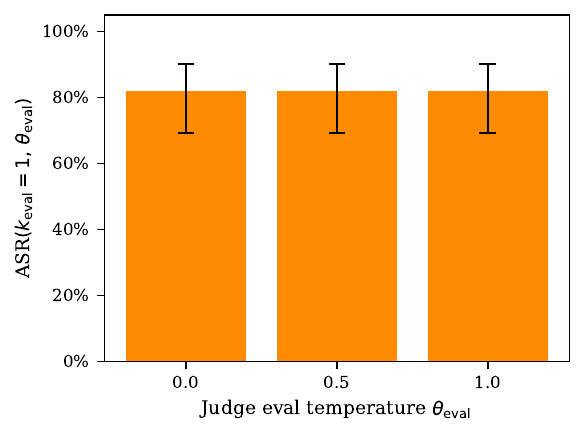} \\
  \end{tabular}
  \caption{\textbf{ASR($k_{eval}=1$, $\theta_{\text{eval}}$)}; Parameters: $\kgen{=}1, \Tgen{=}0.5, \thetagen{=}0.5, \Teval{=}0.5$}
  \label{fig:models_abl3}
\end{figure}
 \clearpage

\subsubsection{Heatmap ASR based on consecutive evaluations at generation}
\begin{figure}[h!]
  \centering
  \begin{tabular}{@{}m{0.3\linewidth}@{\hspace{8pt}}m{0.3\linewidth}@{\hspace{8pt}}m{0.3\linewidth}@{}}
    \multicolumn{1}{c}{\small\textbf{Llama-3.2-1B}} & \multicolumn{1}{c}{\small\textbf{Gemma3-1B}} & \multicolumn{1}{c}{\small\textbf{Granite-3.2-1B}} \\[4pt]
    \includegraphics[width=\linewidth]{assets/figures/abl5_heatmap_kR.pdf} &
    \includegraphics[width=\linewidth]{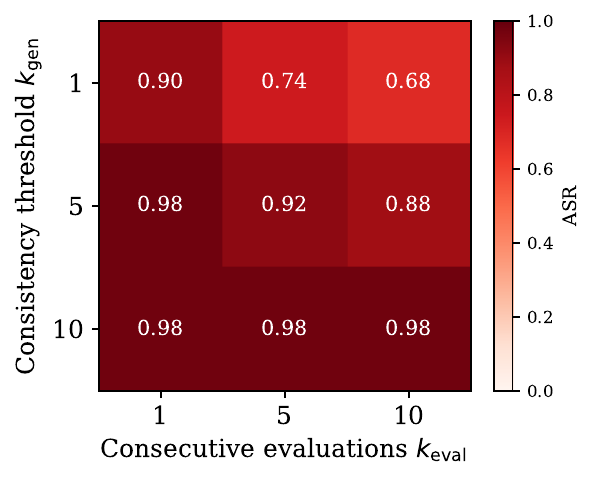} &
    \includegraphics[width=\linewidth]{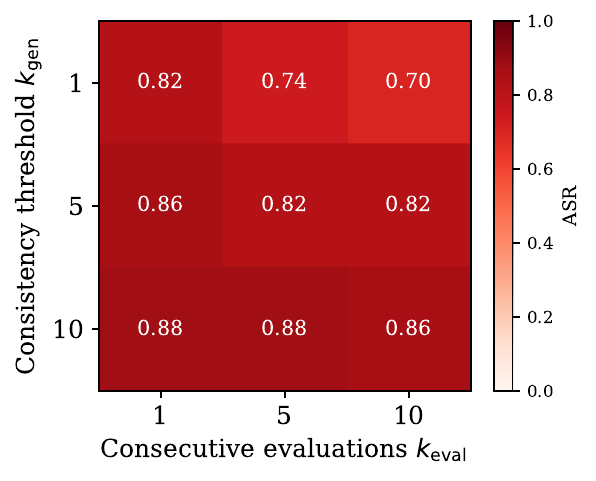} \\
  \end{tabular}
  \caption{Heatmap ASR($k_{eval},k_{gen}$) ; Parameters: $\Tgen{=}0.5, \thetagen{=}0.5, \Teval{=}0.5$}
  \label{fig:models_abl5}
\end{figure}

\subsubsection{Heatmap ASR based on the temperature of the target model at generation}
\begin{figure}[h!]
  \centering
  \begin{tabular}{@{}m{0.3\linewidth}@{\hspace{8pt}}m{0.3\linewidth}@{\hspace{8pt}}m{0.3\linewidth}@{}}
    \multicolumn{1}{c}{\small\textbf{Llama-3.2-1B}} & \multicolumn{1}{c}{\small\textbf{Gemma3-1B}} & \multicolumn{1}{c}{\small\textbf{Granite-3.2-1B}} \\[4pt]
    \includegraphics[width=\linewidth]{assets/figures/abl6_heatmap_kT.pdf} &
    \includegraphics[width=\linewidth]{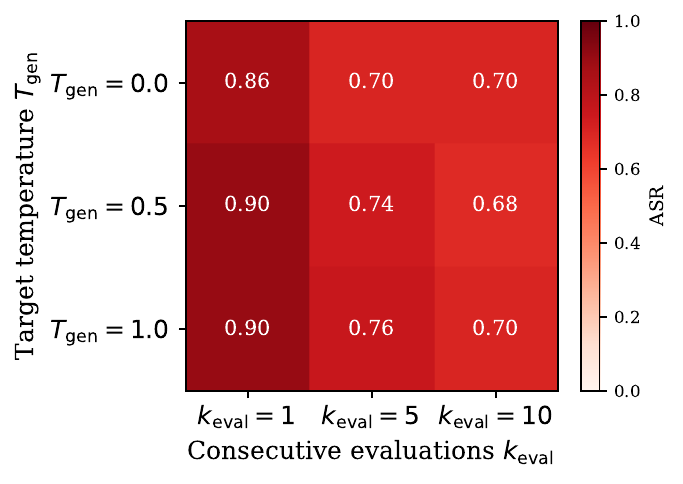} &
    \includegraphics[width=\linewidth]{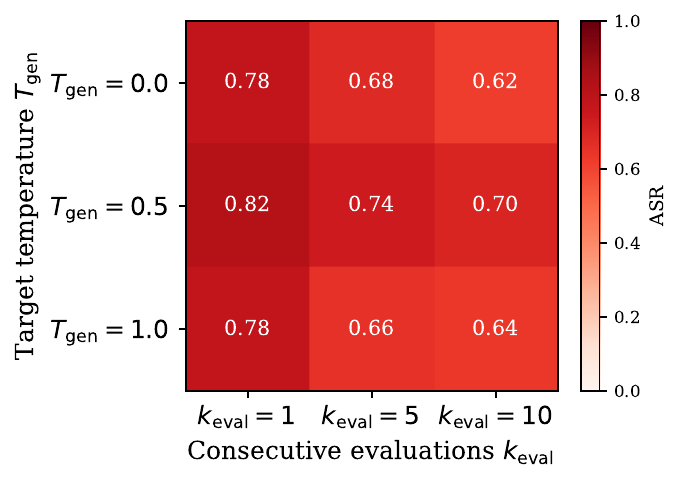} \\
  \end{tabular}
  \caption{Heatmap ASR($k_{eval},\Tgen$) ; Parameters: $\thetagen{=}0.5, k_{gen}=1, \Teval{=}0.5, \thetaeval{=}0.5$}
  \label{fig:models_abl6}
\end{figure}

\subsubsection{Heatmap ASR based on the temperature of the judge at generation}
\begin{figure}[h!]
  \centering
  \begin{tabular}{@{}m{0.3\linewidth}@{\hspace{8pt}}m{0.3\linewidth}@{\hspace{8pt}}m{0.3\linewidth}@{}}
    \multicolumn{1}{c}{\small\textbf{Llama-3.2-1B}} & \multicolumn{1}{c}{\small\textbf{Gemma3-1B}} & \multicolumn{1}{c}{\small\textbf{Granite-3.2-1B}} \\[4pt]
    \includegraphics[width=\linewidth]{assets/figures/abl7_heatmap_ktheta_gen.pdf} &
    \includegraphics[width=\linewidth]{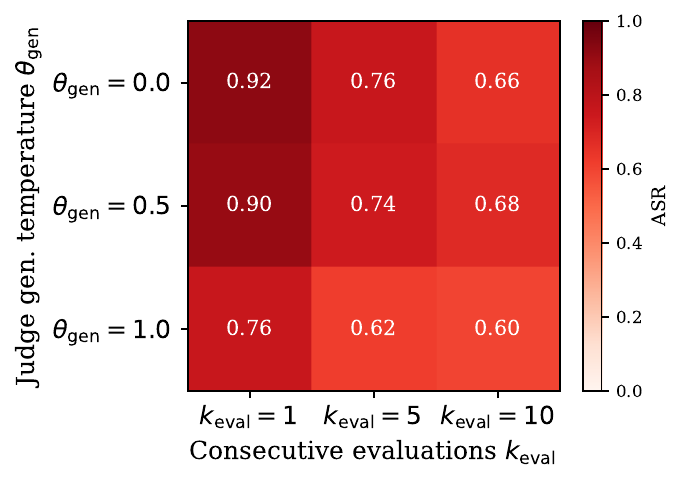} &
    \includegraphics[width=\linewidth]{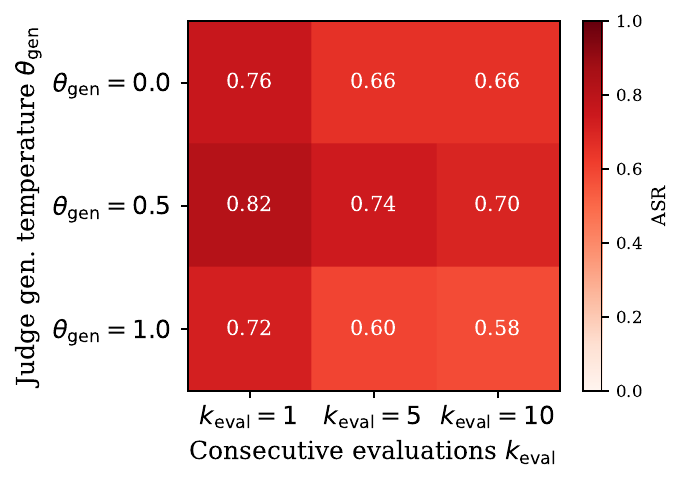} \\
  \end{tabular}
    \caption{Heatmap ASR($k_{eval},\thetagen$) ; Parameters: $\Tgen{=}0.5, k_{gen}=1, \Teval{=}0.5, \thetaeval{=}0.5$}
  \label{fig:models_abl7}
\end{figure}

\clearpage

\subsection{Different attacks}
\label{app:overview-attacks}
\subsubsection{ASR based consecutive judge evaluations}
\begin{figure}[h!]
  \centering
  \begin{tabular}{@{}m{0.455\linewidth}@{\hspace{8pt}}m{0.455\linewidth}@{}}
    \multicolumn{1}{c}{\small\textbf{Best-of-N}} & \multicolumn{1}{c}{\small\textbf{PAIR}} \\[4pt]
      \includegraphics[width=\linewidth]{assets/figures/abl2_asr_k.pdf} &
      \includegraphics[width=\linewidth]{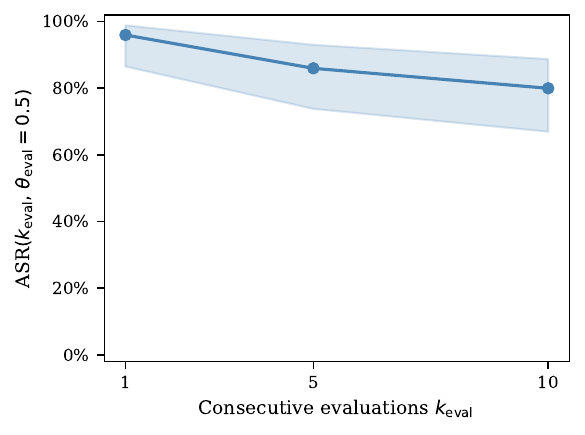} \\[8pt]
    \multicolumn{1}{c}{\small\textbf{TAP}} & \multicolumn{1}{c}{\small\textbf{Crescendo}} \\[4pt]
      \includegraphics[width=\linewidth]{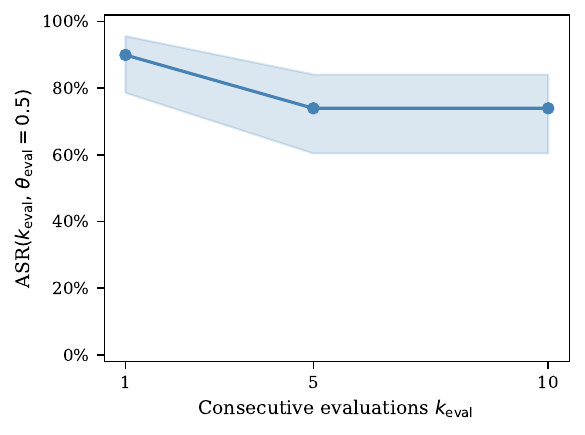} &
      \includegraphics[width=\linewidth]{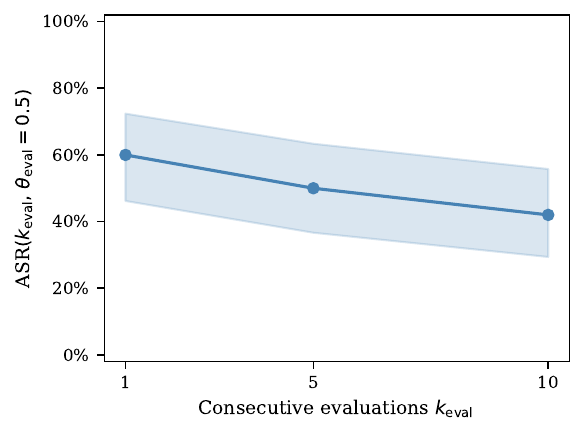} \\
  \end{tabular}
 \caption{\textbf{ASR($k_{eval}$)}; Parameters: $\kgen{=}1, \Tgen{=}0.5, \thetagen{=}0.5, \Teval{=}0.5, \theta_{\text{eval}}{=}0.5$}
  \label{fig:ov_atk_abl2}
\end{figure}

\clearpage
\subsubsection{ASR based on target model temperature at evaluation}
\begin{figure}[h!]
  \centering
  \begin{tabular}{@{}m{0.455\linewidth}@{\hspace{8pt}}m{0.455\linewidth}@{}}
    \multicolumn{1}{c}{\small\textbf{Best-of-N}} & \multicolumn{1}{c}{\small\textbf{PAIR}} \\[4pt]
      \includegraphics[width=\linewidth]{assets/figures/abl1_asr_T.pdf} &
      \includegraphics[width=\linewidth]{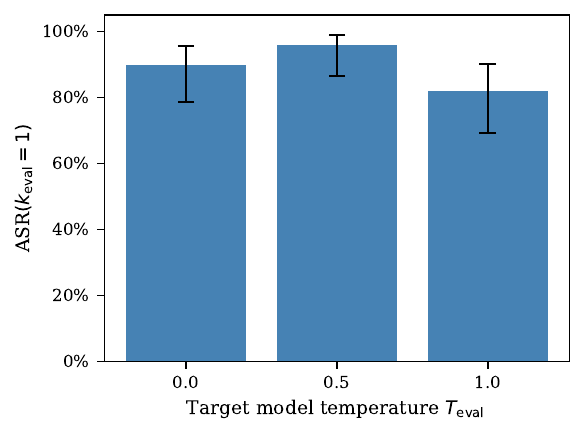} \\[8pt]
    \multicolumn{1}{c}{\small\textbf{TAP}} & \multicolumn{1}{c}{\small\textbf{Crescendo}} \\[4pt]
      \includegraphics[width=\linewidth]{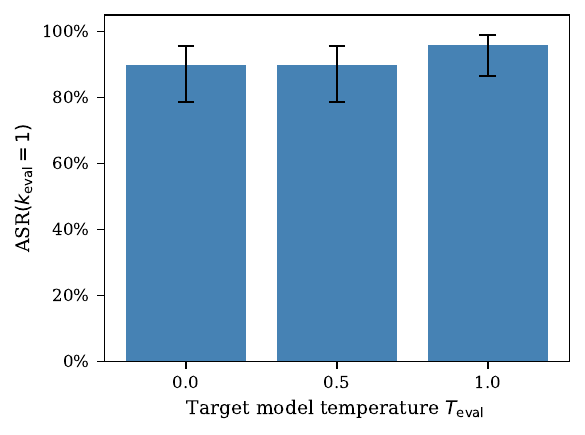} &
      \includegraphics[width=\linewidth]{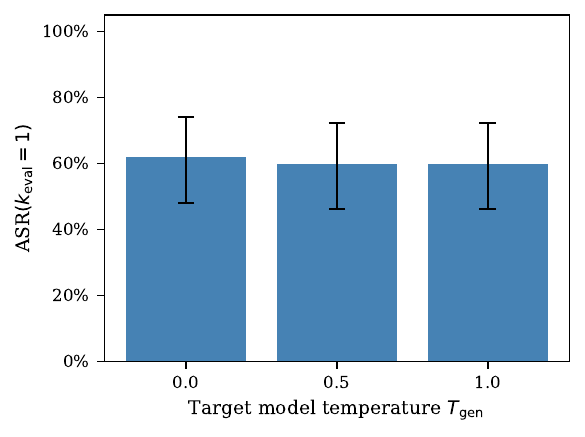} \\
  \end{tabular}
  \caption{\textbf{ASR($k_{eval}=1$, $T_{eval}$)}; Parameters: $\kgen{=}1, \Tgen{=}0.5, \thetagen{=}0.5,\theta_{\text{eval}}{=}0.5$}
  \label{fig:ov_atk_abl1}
\end{figure}

\clearpage

\subsubsection{ASR based on judge model temperature at evaluation}
\begin{figure}[h!]
  \centering
  \begin{tabular}{@{}m{0.455\linewidth}@{\hspace{8pt}}m{0.455\linewidth}@{}}
    \multicolumn{1}{c}{\small\textbf{Best-of-N}} & \multicolumn{1}{c}{\small\textbf{PAIR}} \\[4pt]
      \includegraphics[width=\linewidth]{assets/figures/abl3_asr_theta_eval.pdf} &
      \includegraphics[width=\linewidth]{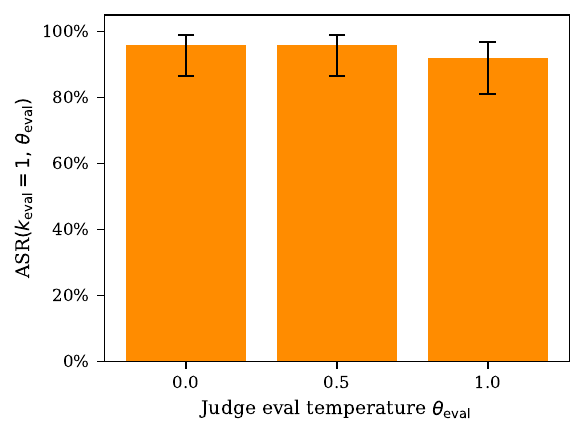} \\[8pt]
    \multicolumn{1}{c}{\small\textbf{TAP}} & \multicolumn{1}{c}{\small\textbf{Crescendo}} \\[4pt]
      \includegraphics[width=\linewidth]{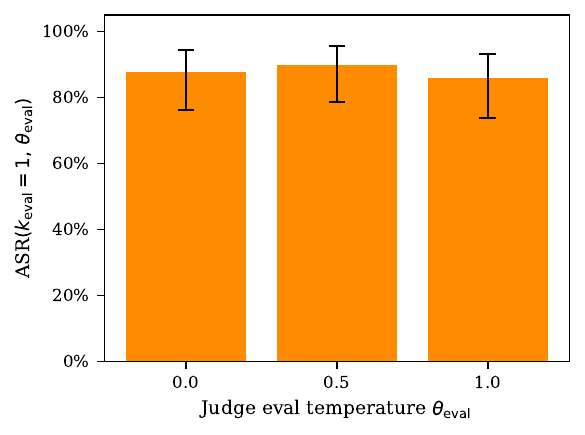} &
      \includegraphics[width=\linewidth]{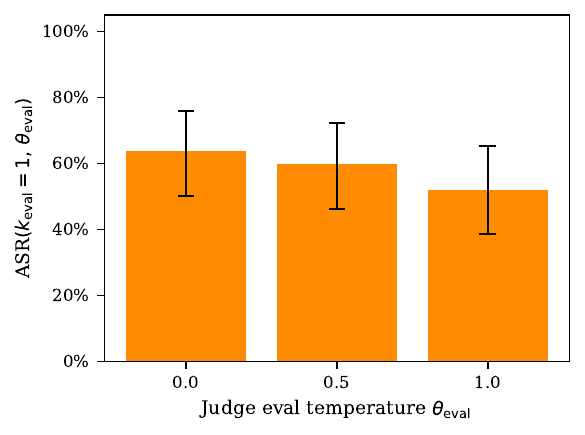} \\
  \end{tabular}
  \caption{\textbf{ASR($k_{eval}=1$, $\theta_{\text{eval}}$)}; Parameters: $\kgen{=}1, \Tgen{=}0.5, \thetagen{=}0.5, \Teval{=}0.5$}
  \label{fig:ov_atk_abl3}
\end{figure}
\clearpage

\subsubsection{Heatmap ASR based on consecutive evaluations at generation}
\begin{figure}[h!]
  \centering
  \begin{tabular}{@{}m{0.455\linewidth}@{\hspace{8pt}}m{0.455\linewidth}@{}}
    \multicolumn{1}{c}{\small\textbf{Best-of-N}} & \multicolumn{1}{c}{\small\textbf{PAIR}} \\[4pt]
      \includegraphics[width=\linewidth]{assets/figures/abl5_heatmap_kR.pdf} &
      \includegraphics[width=\linewidth]{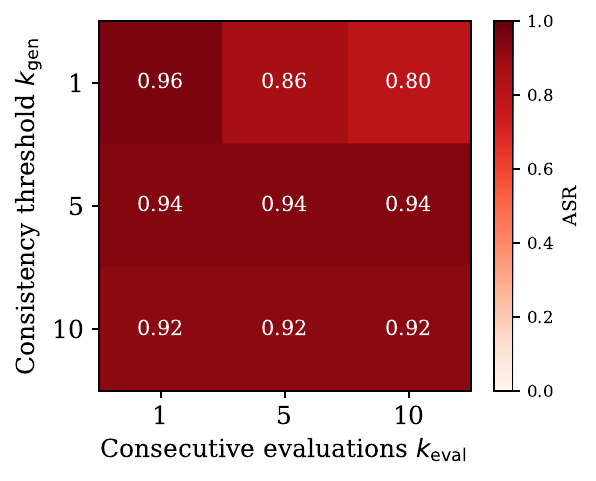} \\[8pt]
    \multicolumn{1}{c}{\small\textbf{TAP}} & \multicolumn{1}{c}{\small\textbf{Crescendo}} \\[4pt]
      \includegraphics[width=\linewidth]{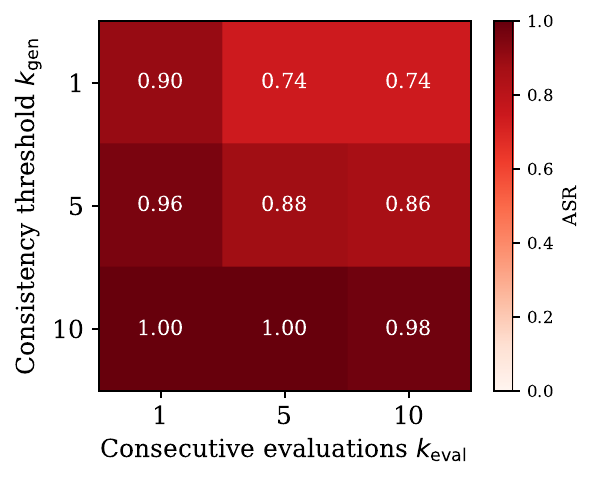} &
      \includegraphics[width=\linewidth]{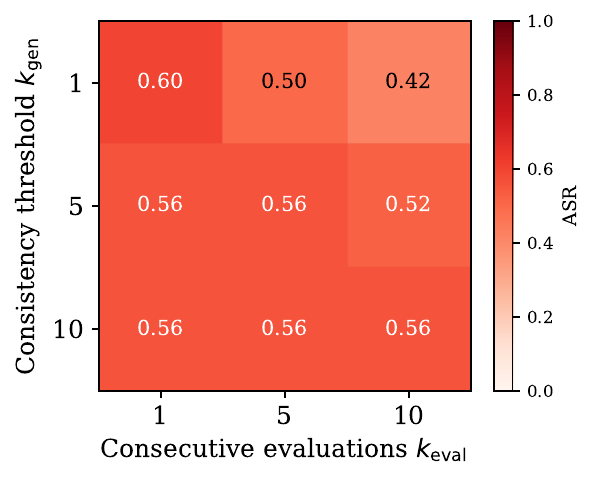} \\
  \end{tabular}
  \caption{Heatmap ASR($k_{eval},k_{gen}$) ; Parameters: $\Tgen{=}0.5, \thetagen{=}0.5, \Teval{=}0.5$}
  \label{fig:ov_atk_abl5}
\end{figure}

\clearpage

\subsubsection{Heatmap ASR based on the temperature of the target model at generation}
\begin{figure}[h!]
  \centering
  \begin{tabular}{@{}m{0.455\linewidth}@{\hspace{8pt}}m{0.455\linewidth}@{}}
    \multicolumn{1}{c}{\small\textbf{Best-of-N}} & \multicolumn{1}{c}{\small\textbf{PAIR}} \\[4pt]
      \includegraphics[width=\linewidth]{assets/figures/abl6_heatmap_kT.pdf} &
      \includegraphics[width=\linewidth]{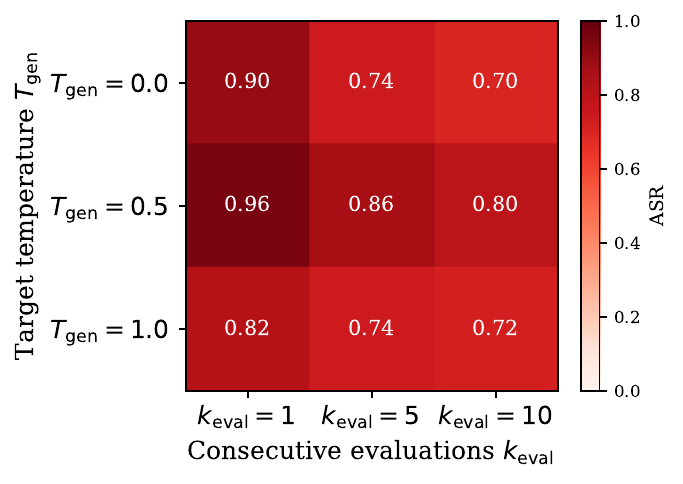} \\[8pt]
    \multicolumn{1}{c}{\small\textbf{TAP}} & \multicolumn{1}{c}{\small\textbf{Crescendo}} \\[4pt]
      \includegraphics[width=\linewidth]{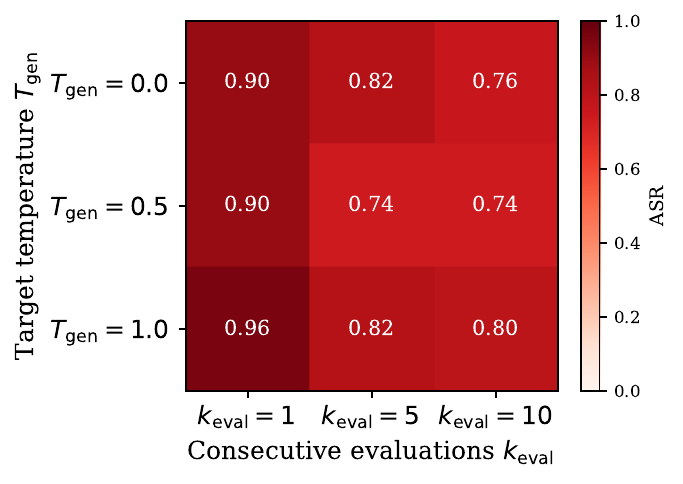} &
      \includegraphics[width=\linewidth]{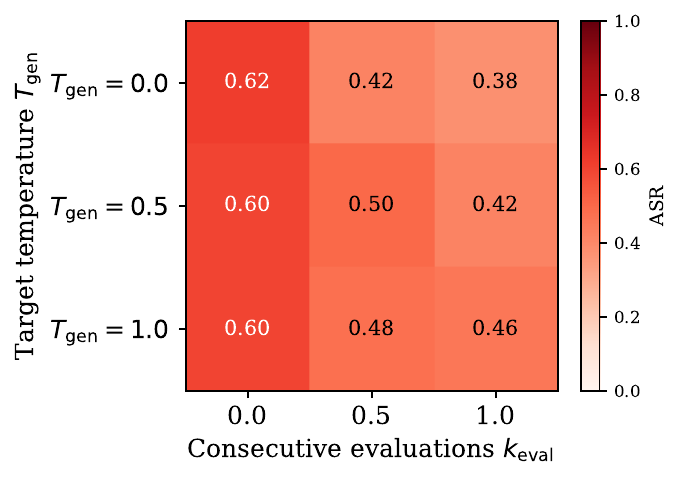} \\
  \end{tabular}
  \caption{Heatmap ASR($k_{eval},\Tgen$) ; Parameters: $\thetagen{=}0.5, k_{gen}=1, \Teval{=}0.5, \thetaeval{=}0.5$}
  \label{fig:ov_atk_abl6}
\end{figure}

\clearpage
\subsubsection{Heatmap ASR based on the temperature of the judge at generation}
\begin{figure}[h!]
  \centering
  \begin{tabular}{@{}m{0.455\linewidth}@{\hspace{8pt}}m{0.455\linewidth}@{}}
    \multicolumn{1}{c}{\small\textbf{Best-of-N}} & \multicolumn{1}{c}{\small\textbf{PAIR}} \\[4pt]
      \includegraphics[width=\linewidth]{assets/figures/abl7_heatmap_ktheta_gen.pdf} &
      \includegraphics[width=\linewidth]{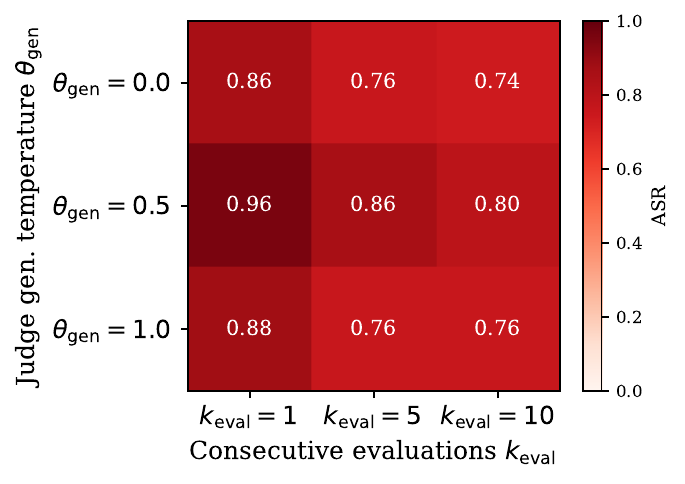} \\[8pt]
    \multicolumn{1}{c}{\small\textbf{TAP}} & \multicolumn{1}{c}{\small\textbf{Crescendo}} \\[4pt]
      \includegraphics[width=\linewidth]{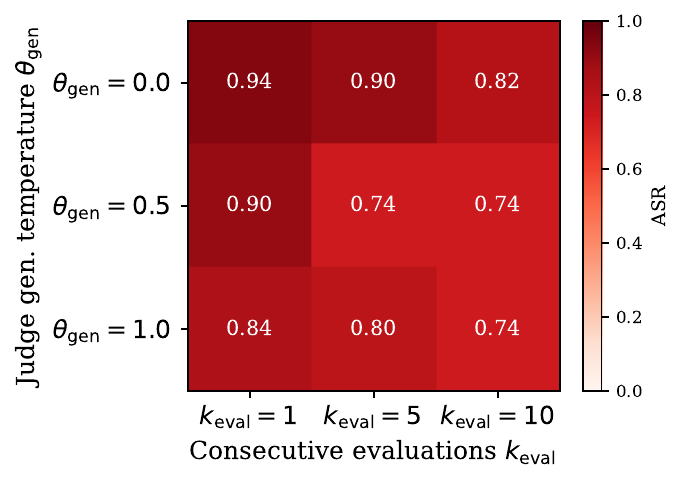} &
      \includegraphics[width=\linewidth]{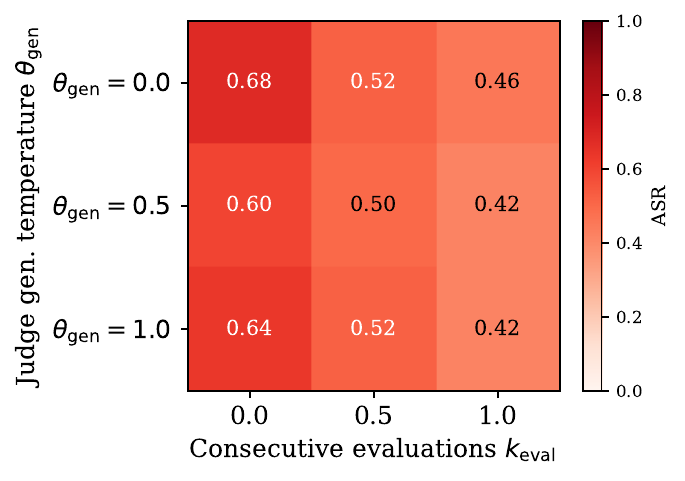} \\
  \end{tabular}
    \caption{Heatmap ASR($k_{eval},\thetagen$) ; Parameters: $\Tgen{=}0.5, k_{gen}=1, \Teval{=}0.5, \thetaeval{=}0.5$}
  \label{fig:ov_atk_abl7}
\end{figure}

\end{document}